\documentclass[a4paper,english,manuscript,superscriptaddress]{revtex4-1}
\usepackage[T1]{fontenc}
\usepackage[latin9]{inputenc}
\usepackage{color}
\usepackage{babel}
\usepackage{float}
\usepackage{amsmath}
\usepackage{amssymb}
\usepackage{graphicx}
\usepackage[unicode=true,pdfusetitle,
 bookmarks=true,bookmarksnumbered=false,bookmarksopen=false,
 breaklinks=false,pdfborder={0 0 0},pdfborderstyle={},backref=false,colorlinks=true]
 {hyperref}
\hypersetup{
 linkcolor=blue,  citecolor=cyan}

\PassOptionsToPackage{normalem}{ulem}
\usepackage{ulem}

\makeatletter

\pdfpageheight\paperheight
\pdfpagewidth\paperwidth

\providecommand{\tabularnewline}{\\}

\usepackage{xcolor}

\makeatother

\begin{document}
\title{Towards a full solution of relativistic Boltzmann equation for quark-gluon matter on GPUs}
\author{Jun-Jie Zhang}
\affiliation{Department of Modern Physics, University of Science and Technology
of China, Hefei 230026, China}
\author{Hong-Zhong Wu}
\affiliation{Department of Modern Physics, University of Science and Technology
of China, Hefei 230026, China}
\author{Shi Pu}
\affiliation{Department of Modern Physics, University of Science and Technology
of China, Hefei 230026, China}
\author{Guang-You Qin}
\affiliation{Institute of Particle Physics and Key Laboratory of Quark and Lepton
Physics (MOE), Central China Normal University, Wuhan, Hubei, 430079, China}
\affiliation{Nuclear Science Division, Lawrence Berkeley National Laboratory, Berkeley,
CA 94720, USA}
\author{Qun Wang}
\affiliation{Department of Modern Physics, University of Science and Technology
of China, Hefei 230026, China}
\begin{abstract}

We have developed a  numerical framework for a full solution of the relativistic Boltzmann equations for the quark-gluon matter using the multiple Graphics Processing Units (GPUs) on distributed clusters. Including all the $2 \to 2$ scattering processes of 3-flavor quarks and gluons, we compute the time evolution of distribution functions in both coordinate and momentum spaces for the cases of pure gluons, quarks and the mixture of quarks and gluons. By introducing a symmetrical sampling method on GPUs which ensures the particle number conservation, our framework is able to perform the space-time evolution of quark-gluon system towards thermal equilibrium with high performance. We also observe that the gluons naturally accumulate in the soft region at the early time, which may indicate the gluon condensation.

\end{abstract}
\maketitle

\section{\label{sec:introduction}introduction}


Relativistic Boltzmann equation (BE), an effective theory of many-body systems, is a profound and widely used tool to study the properties of the systems out of equilibrium or in thermal equilibrium.
Recently, BE is often applied to study the problem of early thermalization, which remains to be one of the \emph{``greatest unsolved problems}\emph{''} \citep{Fukushima2016} in relativistic heavy-ion collisions, which collide two accelerated nuclei to create a hot and dense deconfined nuclear matter,  named quark-gluon plasma (QGP).
The space-time evolution of QGP has been well described by relativistic hydrodynamics simulations.
The success of hydrodynamical models on soft hadron production and collective flows provides strong evidence for the rapid thermalization of the quark-gluon system to create a strongly-interacting QGP \citep{Pang2018,Heinz2002a,Heinz2003,Heinz2002,Mueller2007}.
The time scale expected for thermalization is estimated to be less than $1\textrm{fm/c}$ \citep{Heinz2003} or even shorter than
$0.25\textrm{fm/c}$ \citep{Bozek2011,Broniowski2009} in nucleus-nucleus collisions at Relativistic Heavy-Ion Collider (RHIC) and the Large Hadron Collider (LHC).
However, it remains to be a puzzle how an over-occupied gluonic system with weak coupling can reach thermal equilibrium within such a short time scale \citep{Heinz2002}.

\subsection{Background}

The study of the systems under the framework of BE with suitable initial conditions, also referred to as the kinetic approach, is a well-established method for probing the real-time quark and gluon dynamics in the dilute regime at weakly coupling limit \citep{Arnold2003a,Chen2013,Mueller2000a}.
However, a full solution of the relativistic BE involving all parton species, e.g., $u,d,s$ quarks, their antiparticles, and gluons, is still challenging both analytically and numerically due to the complexity of the collision integral, higher dimensions and computing resources.


The typical initial condition for relativistic heavy-ion collisions \citep{Itakura2007} is an overpopulated gluonic state named Color Glass Condensate (CGC)~\citep{Mueller2000,Blaizot1987,Gribov1983,Eskola2000,Jalilian-Marian1997,McLerran1994,Krasnitz2001,Kovner1995a}, which is formed in the dynamical balance between the splitting and fusion of gluons in the small-$x$ region.
The occupation number of small-$x$ gluons is of order $1/\alpha_{s}$ \citep{Gelis2010}.
The gluon number grows until the gluon size is larger than $1/Q_{s}$ \citep{McLerran2008}, with $Q_{s}$ being the saturation scale. After the CGC state, the glasma, a state of color electromagnetic fields, may be formed \citep{Lappi2006,Lappi2003,Krasnitz2000,Kovner1995,Gelis2012,McLerran2008}.
Currently, how the glasma transits to a thermalized QGP in a short time scale is not well understood yet, which is often referred to as the early thermalization puzzle.

There have been various studies focusing on the evolution of the glasma stage.
For example, the ``bottom-up scenario'' \citep{Baier2001} estimates a thermalization time of order $\tau_{th}\sim\alpha_{s}^{-13/5}/Q_{s}$, which is unfortunately too large compared to the time scale required by the hydrodynamics.
To reconcile this discrepancy, many other possible mechanisms have been considered.
One interesting mechanism is the so-called plasma instability \citep{Mrowczynski1988,Mrowczynski1993,Mrowczynski1994,Mrowczynski1997,Randrup2003,Weibel1959,Buneman1958}. It originates from the anisotropic momentum distribution in the plasma and may drastically speed up the process of glasma equilibration \citep{Arnold2003,Strickland2006,Romatschke2004}. However, some studies also imply that the plasma instability may not play a significant role at the early stage \citep{Berges2015,Arnold2004}, and a scaling solution may also be required \citep{Mueller2006}.
Another mechanism is the Bose-Einstein condensation of gluons.
It has been suggested that the gluon condensation at the early stage \citep{Scardina2014,Blaizot2013} may accelerate the thermalization process \citep{Blaizot2014,Huang2015,Zhou2017,Xu2013,Xu2013a}.
The influence of fermions and masses in forming condensation has also been discussed in Refs. \citep{Xiao-Ming2005,Biro1982,Kurkela2019a,Kurkela2019,Blaizot2016}.
However, it is argued that the inelastic scattering processes may strongly hinder the effect of gluon condensation \citep{Blaizot2017,Lenkiewicz2019}.
In fact, the role of inelastic scatterings in the thermalization is still not quite clear so far.
It has been suggested long time ago that the inelastic processes might be essential for thermalization \citep{Wong1996}.  
Some works by solving BE
with the test particle method including $2\rightarrow3$ processes
has obtained the results close to the ``bottom-up'' scenario \citep{Shin2003}.
Another simulation from the Boltzmann approach of multiparton scattering (BAMPS),
a package for solving BE using test particle method,
suggests that the 
 bremsstrahlung from $2\rightarrow3$ processes increases the efficiency of thermalization \citep{Xu2006}.
Later studies from BAMPS imply that the inverse processes, i.e. the $3\rightarrow2$ processes, inhibit the $2\rightarrow3$ processes. With both processes included,
the time scale of thermal equilibration from the BAMPS is of order $\alpha_{s}^{-2}\ln(\alpha_{s})^{-2}Q_{s}^{-1}$ \citep{El2008}.
It has also been argued that that the inclusion of all next-to-leading order processes may make
the equilibration considerably faster than the simple $2\rightarrow3$
processes \citep{Xiong1994}.

\subsection{Motivation}

Some of the above studies require solving relativistic BE numerically.
Historically, a full numerical solution of the non-relativistic BE has always been a challenge due to its
high dimensions and the intrinsic physical properties \citep{Dimarco2014,Romatschke2011,Jaiswal2019}.
Even in today\textquoteright s petascale clusters, BE still presents
a substantial computational challenge \citep{Jaiswal2019}.
In the real application of non-relativistic BE, one usually needs very dense
spatial grids to describe complicated effects related to pressure,
temperature, and turbulence, etc. 

The main difference between the relativistic and non-relativistic BE lies in the collision term and
the coupled equations. In the relativistic BE, the collision integrals
are usually much more complicated than those in the non-relativistic case.
For example, in our relativistic BE, there are seven particle species with complicated
scattering matrix, while in the non-relativistic case, one usually
deals with single species of particles in a gas or liquid. Despite
the heavy workload of the collision term, our relativistic BE also
contains coupled equations of seven species, which need to be solved
simultaneously. Furthermore, the distributions for fermions
should not exceed unity due to Pauli exclusion principle. This limitation requires
that the time step of the evolution be sufficiently small,
which in turn drastically slows down the speed of the code.

Though complicated, several numerical tools based on the parton cascade model
\citep{Geiger1995,Geiger1992} have been developed in the market,
e.g., BAMPS \citep{Xu2005} and ZPC \citep{Zhang1998,Zhang1998b}.
Another way is the lattice Boltzmann approach \citep{Chen1998,Succi2015,Dimarco2016},
which has been utilized as a fast lattice Boltzmann solver for relativistic
hydrodynamics with relaxation time approximation \citep{Romatschke2011,Mendoza2010a,Hupp2011}.
In addition, the straightforward implementations of effective
kinetic theory (e.g. see Ref.~\citep{Arnold2003a} for the
theoretical framework) for pure gluons~\citep{Kurkela2014,Kurkela2015}
and quark-gluon systems~\citep{Kurkela2019,Kurkela2019a,Keegan2016}
also provide us physical insights for pre-thermalization.
While these models and approaches have succeeded in describing the non-equilibrium evolution of quark-gluon matter
with a set of parameters given by the physical consideration for simplicity, a full solution of the BE is still demanded for a comprehensive understanding of the thermalization puzzle.

\subsection{Our numerical framework}


In this study, we develop a numerical framework for the full numerical solutions of the relativistic BE with the help of the state-of-art GPUs.
GPU, known for its high clock rate, high instruction per cycle \citep{Gorelick2014},
and multiple cores, makes more and more contributions to computational physics nowadays \citep{Nickolls2010}. 
Some calculations, which are
extremely difficult in the old-time, are now within the scope \citep{Bauce2014,Clark2010,Gutierrez2010,Bazow2016,Brechtken2012,Seiskari2012,Clark2016,Winter2014,Zhang2019a}.
Some physical phenomena could be understood by the simulations via
GPUs \citep{Adler2017}. With the new GPU techniques, various attempts
have been done to tackle the BE from different aspects for non-relativistic
cases \citep{Colmenares2013,Frezzotti2011,Kloss2010,Januszewski2014,Jaiswal2019,Colmenares2014}.
The GPU techniques also motivate us to develop the framework for solving the relativistic BE in the context of relativistic heavy-ion collisions, despite a series of methodological challenges that have no counterparts in the non-relativistic realm.
As a first step, we only consider the $2\rightarrow2$ scattering processes in the current work.

The difficulties in solving the relativistic BE originate from two aspects. First, the collision terms are high dimensional
integrals. In this work, we use the package ZMCintegral 5.0~\citep{Wu:2019tsf,Zhang2019} to perform these high dimensional collisional integrals. ZMCintegral is an open-source Python package developed by some of the authors
in this work. The second difficulty is the issue of particle number non-conservation
due to the discrete Monte Carlo (MC) integration, see also Refs. \citep{Epelbaum:2014mfa,Li:2012qt}.
To achieve a strict particle number conservation in the CPU framework
will usually cost lots of computing time. In our work, we propose a ``symmetrical
sampling'' method for the collisional integrals via GPUs. With the
help of new features of GPUs, named CUDA atomic operations \citep{cudaAutomicOperation,cudaAutomicOperation2},
we can achieve the strict particle number conservation with acceptable computing time.

Our numerical framework provides a full solution of the BE
with complete $2\rightarrow2$ scattering processes. The program is developed
with the combination of Python library Numba \citep{Lam2015} and
Ray \citep{PMoritz2018}, which enable the manipulation of GPU devices
on distributed clusters.
We will test the code in many aspects, for instance,
the stability of collisional integrals, the particle number conservation, and the total
energy conservation. We will also show the time evolution
of the distribution functions in both coordinate and momentum spaces.

The structure of this paper is as follows. In Sec. \ref{sec:boltzmann-equation-qgp},
we briefly review the ordinary BE for thermal quarks and gluons. Next
we introduce our numerical framework in Sec. \ref{sec:numerical-method-involved}.
We first discuss how to get stable numerical results in collision integral
in subsec. \ref{subsec:Delta-function-and} and then introduce the method
to keep the particle numbers conserved in subsec \ref{subsec:Particle-number-conservation}.
We will present our numerical results in Sec. \ref{sec:performance}.
In subsec. \ref{subsec:Stability-test-of} and \ref{subsec:Test-of-Particle},
we test the stability of the collisional integrals and the particle number
conservation. We discuss the total energy conservation in subsec. \ref{subsec:Total-energy-conservation}.
With some physical initial conditions, we show the time evolution
of the system in both coordinate and momentum spaces in subsec. \ref{subsec:Evolution-of-coordinates}
and \ref{subsec:Evolution-in-momentum}, respectively. The summary of our paper will be presented in Sec \ref{sec:Conclusion}.

Throughout this work, we choose the metric $g^{\mu\nu}=\textrm{diag}\{+,-,-,-\}$
and the space and momentum four vectors as $x^{\mu}=(t,\mathbf{x})$
and $p^{\mu}=(E_{p},\mathbf{p})$. For a momentum $k_{a}^{\mu}$,
we choose $\mu=(t,x,y,z)$ for the space-time index and $a=1,2,3$
for $u,d,s$ flavors.

\section{Boltzmann equation for quark gluon matter \label{sec:boltzmann-equation-qgp}}

In this section, we will briefly review the ordinary BE for the thermal
quarks and gluons in the leading-log order. More details can be found in our previous systematic studies in Ref. \citep{Chen2013,Chen:2013tra}.

The relativistic BE, which is an effective theory for relativistic
many-body systems, describes the evolution of system in the phase
space. The general expression for the BE reads,
\begin{equation}
\frac{d}{dt}f_{p}(t,\mathbf{x},\mathbf{p})\equiv\frac{\partial}{\partial t}f_{p}+\frac{\partial\mathbf{x}}{\partial t}\cdot\nabla_{x}f_{p}+\frac{\partial\mathbf{p}}{\partial t}\cdot\nabla_{p}f_{p}=C[f_{p}],
\end{equation}
where $f_{p}(t,\mathbf{x},\mathbf{p})$ is the distribution function
and $C[f]$ is called the collision term. The $\partial\mathbf{x}/\partial t$
and $\partial\mathbf{p}/\partial t$ are the effective velocity and
effective force for the particles, respectively. One can derive the
$\partial\mathbf{x}/\partial t$ and $\partial\mathbf{p}/\partial t$
from the equation of motion of the action for a single particle. In
our study, the action in the classical level, i.e. up to the order
of $\hbar^{0}$, is $S=\int dt(\mathbf{p}\cdot\frac{d\mathbf{x}}{dt}-E_{p})$,
with $E_{p}$ being the particle's energy. For simplicity, we neglect
the particle's physical mass, but the non-vanishing thermal mass $m(x)$
depends on the space-time in general. Accordingly, the particle's
energy $E_{p}(x)\equiv\sqrt{\mathbf{p}^{2}+m^{2}(x)}$ depends
on the space-time.

For thermal quark-gluon matter, the BE has the following general
structure \citep{Jeon1995,Arnold2003a,Chen2013}:
\begin{equation}
\frac{\partial f_{p}^{a}(t,\mathbf{x},\mathbf{p})}{\partial t}+\frac{\mathbf{p}}{E_{p}^{a}}\cdot\nabla_{x}f_{p}^{a}(t,\mathbf{x},\mathbf{p})-\nabla_{x}E_{p}^{a}\cdot\nabla_{p}f_{p}^{a}(t,\mathbf{x},\mathbf{p})=\mathcal{C}_{a},\label{eq:boltzmann equation general form}
\end{equation}
where $f_{p}^{a}(t,\mathbf{x},\mathbf{p})$ denotes the color and
spin averaged distribution function for particle $a$, and $a=q,\bar{q},g$
stands for quarks, anti-quarks and gluons. $E_{p}^{a}(\mathbf{x})=\sqrt{\mathbf{p}^{2}+m_{a}^{2}(\mathbf{x})}$
and $C_{a}$ are the energy and collision term for particle $a$,
respectively. $-\nabla_{x}E_{p}^{a}$ is an effective force, which comes
from the equation of motion of $\partial\mathbf{p}/\partial t$ \citep{Jeon1995,Arnold2003a,Chen2013}.

In the present work,  we only consider the $2\rightarrow2$ scatterings.
The collision term for a quark of flavor $a$ can be obtained,
\begin{eqnarray}
N_{q}\mathcal{C}_{q_{a}} & = & \frac{1}{2}C_{q_{a}q_{a}\leftrightarrow q_{a}q_{a}}+C_{q_{a}\bar{q}_{a}\leftrightarrow q_{a}\bar{q}_{a}}+\frac{1}{2}C_{gg\leftrightarrow q_{a}\bar{q}_{a}}+C_{q_{a}g\leftrightarrow q_{a}g}\nonumber \\
 &  & +\sum\limits _{b,b\neq a}(C_{q_{a}q_{b}\leftrightarrow q_{a}q_{b}}+C_{q_{a}\bar{q}_{b}\leftrightarrow q_{a}\bar{q}_{b}}+C_{q_{b}\bar{q}_{b}\leftrightarrow q_{a}\bar{q}_{a}}),\label{eq:C_quark}
\end{eqnarray}
where $N_{q}=2\times3=6$ is the quark helicity and color degeneracy
factor and the factor $1/2$ is included when the initial state is
composed of two identical particles. For a gluon, the collision term
reads
\begin{equation}
N_{g}\mathcal{C}_{g}=\frac{1}{2}C_{gg\leftrightarrow gg}+\sum\limits _{a}(C_{gq_{a}\leftrightarrow gq_{a}}+C_{g\bar{q}_{a}\leftrightarrow g\bar{q}_{a}}+C_{q_{a}\bar{q}_{a}\leftrightarrow gg}),\label{eq:C_gluon}
\end{equation}
where $N_{g}=2\times8=16$ is the gluon helicity and color degeneracy
factor.

The collision term for $2\rightarrow2$ scatterings, $a(k_{1})+b(k_{2})\rightarrow c(k_{3})+d(p)$,
has the following general expression,
\begin{eqnarray}
C_{ab\rightarrow cd} & \equiv & \int\prod_{i=1}^{3}\frac{d^{3}k_{i}}{(2\pi)^{3}2E_{k_{i}}}\frac{|M_{ab\leftrightarrow cd}|^{2}}{2E_{p}}(2\pi)^{4}\delta^{(4)}(k_{1}+k_{2}-k_{3}-p)[f_{k_{1}}^{a}f_{k_{2}}^{b}F_{k_{3}}^{c}F_{p}^{d}-F_{k_{1}}^{a}F_{k_{2}}^{b}f_{k_{3}}^{c}f_{p}^{d}],\nonumber \\
\label{eq:C_a_b}
\end{eqnarray}
where $F_{p}^{g}=1+f_{p}^{g},F_{p}^{q(\bar{q})}=1-f_{p}^{q(\bar{q})}$
and $M_{ab\rightarrow cd}$ is the matrix element in which all colors
and helicities of the initial and final states are summed over. We
summarize all $2\rightarrow2$ scattering matrix elements in Table \ref{tab:Matrix-elements-squared}
in the Appendix \ref{matrix element}.
In our numerical calculation, the tree-level matrix
elements for all $2\leftrightarrow2$ scattering processes are set as the
default configuration. We also provide an application programming interface (API) for users
to define their own matrix elements for some specific purposes.

When the system reaches the global thermal equilibrium, the distribution
functions should satisfy the ordinary Bose-Einstein or Fermi-Dirac
distributions,
\begin{eqnarray}
f_{p}^{g} & = & \frac{1}{e^{(E_{p}-\mu_{g})/T}-1},\label{eq:fg}\\
f_{p}^{q(\bar{q})} & = & \frac{1}{e^{(E_{p}\mp\mu_{q})/T}+1},\label{eq:fq}
\end{eqnarray}
where $T$ is the temperature and $\mu_{a}$ is the chemical potential
for particle $a$.

The thermal masses of gluon and quark (anti-quark) are usually written
as \citep{Jeon1995,Chen2013},
\begin{eqnarray}
m_{g}^{2}(x) & = & \frac{2g^{2}}{d_{A}}\int\frac{d^{3}p}{(2\pi)^{3}2E_{p}}\Big[N_{g}C_{A}f_{p}^{g}+\sum_{i=1}^{N_{f}}N_{q_{i}}C_{F}(f_{p}^{q_{i}}+f_{p}^{\bar{q}_{i}})\Big],\label{eq:mg}\\
m_{q_{i}}^{2}(x) & = & m_{\bar{q}_{i}}^{2}(x)=2C_{F}g^{2}\int\frac{d^{3}p}{(2\pi)^{3}2E_{p}}(2f_{p}^{g}+f_{p}^{q}+f_{p}^{\bar{q}}),\label{eq:mq}
\end{eqnarray}
where $N_{f}$ is taken to be 3 a we only consider $q_{i}=u,d,s$
quarks and their anti-particles. Note that, when adding the contribution
of $1\rightarrow2$ scattering, the thermal mass can be introduced
in a systematic way, and the invariant momentum differential piece
$d^{3}p/[(2\pi)^{3}E_{p}]$ could be replaced by $d^{3}p/[(2\pi)^{3}|\mathbf{p}|]$
in all momentum integrals, see e.g. in \citep{Arnold2003a,Kurkela2019,Kurkela2019a}.
In the first time step of evolution, since we have no prior information of particle
masses, we will use $E_{p}=|\mathbf{p}|$ to perform the calculation
in Eq. (\ref{eq:mg}) and (\ref{eq:mq}). In later time steps, we
will use the normal $E_{p}^{a}(x)=\sqrt{\mathbf{p}^{2}+m_{a}^{2}(x)}$.
This iteration approach is reasonable since the difference $|E_{p}-p|$
from non-zero masses is of higher order \citep{Chen2013}.

In our calculations, we need to check the total particle number conservation.
The particle number is defined as:
\begin{eqnarray}
N(g) & = & \int d^{3}x\frac{d^{3}p}{(2\pi)^{3}}f_{g}N_{g}\nonumber \\
N(q_{i}) & = & \int d^{3}x\frac{d^{3}p}{(2\pi)^{3}}f^{q_{i}}N_{q_{i}}.\label{eq:particle numbers}
\end{eqnarray}
Note that, in general, the total number for each type of particles (e.g.
gluons, quarks and anti-quarks) is not conserved due to the strong
interaction. However, since we only consider $2\rightarrow2$
scatterings in this work, the total particle number $N_{g}+\sum_{i=1}^{N_{f}}(N_{q_{i}}+N_{\bar{q}_{i}})$ is conserved.
We will check and confirm the total particle number conservation at each time step of our numerical simulations.

Since the thermal masses of particles depend on the space and time, the ordinary kinetic energy-momentum tensor,
\begin{eqnarray}
T_{kin}^{\mu\nu}(x) & = & \sum_{a}\int\frac{d^{3}\mathbf{p}}{(2\pi)^{3}E_{p}^{a}}N_{a}p^{\mu}p^{\nu}f_{p}^{a}(x),\label{eq:energy momentum tensor}
\end{eqnarray}
is not conserved \citep{Jeon1995}. Instead, we have:
\begin{eqnarray}
\partial_{\mu}T_{kin}^{\mu\nu}(x) & = & S_{ex}^{\nu}(x)=\frac{1}{2}\sum_{a}\partial^{\nu}m_{a}^{2}\int\frac{d^{3}\mathbf{p}}{(2\pi)^{3}E_{p}^{a}}N_{a}f_{p}^{a},\label{eq:current conservation}
\end{eqnarray}
where $S_{ex}^{\nu}(x)$ is a source term due to the mass variations.

\section{numerical framework for solving Boltzmann equation \label{sec:numerical-method-involved}}

In order to keep the total particle number conserved, we first rewrite the BE in Eq. (\ref{eq:boltzmann equation general form})
as follows,
\begin{equation}
\frac{\partial f_{p}^{a}(x)}{\partial t}+\nabla_{x}\cdot\left[\frac{\mathbf{p}}{E_{p}^{a}}f_{p}^{a}(x)\right]-\nabla_{p}\cdot\left[(\nabla_{x}E_{p}^{a})f_{p}^{a}(x)\right]=C_{a}[f],\label{eq:left handside}
\end{equation}
where we have used the following identity,
\begin{equation}
-\left[\nabla_{x}\cdot\frac{\mathbf{p}}{E_{p}^{a}}\right]+\left[\nabla_{p}\cdot(\nabla_{x}E_{p}^{a})\right]=0.\label{eq:cancellation of diff terms}
\end{equation}
The form of Eq. (\ref{eq:left handside}) ensures the conservation of total particle numbers when periodical boundary conditions are applied in phase space.
Then using central difference, we can express the left-hand side of Eq. (\ref{eq:left handside})
into discrete form as follows,
\begin{align}
 & \frac{f_{p}^{a}(x+\Delta t)-f_{p}^{a}(x)}{\triangle t}+\sum_{i=1,2,3}\left\{ \frac{1}{2\triangle x_{i}}\left[\frac{p_{i}f_{p}^{a}(x_{i}+\triangle x_{i})}{E_{p}^{a}(x_{i}+\triangle x_{i})}-\frac{p_{i}f_{p}^{a}(x_{i}-\triangle x_{i})}{E_{p}^{a}(x_{i}-\triangle x_{i})}\right]\right.\nonumber \\
 & -\frac{1}{2\triangle p_{i}}\left[\frac{f_{p_{i}+\triangle p_{i}}^{a}(x)}{2\triangle x_{i}}\left(E_{p_{i}+\triangle p_{i}}^{a}(x_{i}+\triangle x_{i})-E_{p_{i}+\triangle p_{i}}^{a}(x_{i}-\triangle x_{i})\right)\right.\nonumber \\
 & \left.\left.-\frac{f_{p_{i}-\triangle p_{i}}^{a}(x)}{2\triangle x_{i}}\left(E_{p_{i}-\triangle p_{i}}^{a}(x_{i}+\triangle x_{i})-E_{p_{i}-\triangle p_{i}}^{a}(x_{i}-\triangle x_{i})\right)\right]\right\} .\label{eq:descrete left hand}
\end{align}

\subsection{The $\delta$-function and the collision term \label{subsec:Delta-function-and}}

Now we look at the collision term in the right-hand side of Eq. (\ref{eq:boltzmann equation general form}) or Eq. (\ref{eq:left handside}).
Usually, one can integrate over the momentum $d^{3}k_{i}$ with the $\delta$-function,
\begin{equation}
\delta^{(4)}(k_{1}+k_{2}-k_{3}-p)=\delta^{(3)}(\mathbf{k}_{1}+\mathbf{k}_{2}-\mathbf{k}_{3}-\mathbf{p})\delta(E_{1}+E_{2}-E_{3}-E_{p}),\label{eq:delta function}
\end{equation}
which can reduce the number of integral variables. Here, we have two
choices: either expressing $\mathbf{k}_{2}$ by $\mathbf{k}_{3}+\mathbf{p}-\mathbf{k}_{1}$
or expressing $\mathbf{k}_{3}$ by $\mathbf{k}_{1}+\mathbf{k}_{2}-\mathbf{p}$.
These two choices can lead to different numerical behaviors.
In this work, we choose the first choice which will make our numerical integrations more stable than the second one (as we will show later).

With the help of the $\delta$-function, we can integrate over $d^{3}\mathbf{k}_{2}$ and obtain,
\begin{align}
\int\prod_{i=1}^{3}\frac{d^{3}\mathbf{k}_{i}}{(2\pi)^{3}2E_{k_{i}}}\delta^{(4)}(k_{1}+k_{2}-k_{3}-p) & =\int\frac{d^{3}\mathbf{k}_{1}}{(2\pi)^{3}2E_{1}}\frac{d^{3}\mathbf{k}_{3}}{(2\pi)^{3}2E_{3}}\frac{1}{(2\pi)^{3}2E_{2}}\delta(E_{1}+E_{2}-E_{3}-E_{p})\nonumber \\
 & =\frac{1}{(2\pi)^{9}}\int\frac{d^{3}\mathbf{k}_{3}dk_{1}^{x}dk_{1}^{y}}{2E_{1}2E_{2}2E_{3}}\sum_{i=\pm}\frac{1}{|J(k_{1z}^{i})|},\label{eq:work out delta E}
\end{align}
where we have used
\begin{eqnarray}
\delta(E_{1}+E_{2}-E_{3}-E_{p}) & = & \sum_{i=\pm}\frac{1}{|J(k_{1z}^{i})|}\delta(k_{1z}-k_{1z}^{i}),\label{eq:delta replace}
\end{eqnarray}
with
\begin{eqnarray}
J(k_{1z}^{\pm}) & = & \frac{k_{1z}^{\pm}}{E_{1}}-\frac{-k_{1z}^{\pm}+k_{3z}+p_{z}}{E_{2}},\nonumber \\
k_{1z}^{\pm} & = & \mathrm{Root}[E_{1}+E_{2}-E_{3}-E_{p}=0].
\end{eqnarray}
There are two roots for $k_{1z}$ from the equation $E_{1}+E_{2}-E_{3}-E_{p}=0$, and $k_{1z}$ has the form of $k_{1z}^{\pm}\equiv\frac{A\pm\sqrt{H}}{B}$,
where $A,B,H$ are functions of $k_{1}^{x},k_{1}^{y}$ and $\mathbf{k}_{3}$.

Substituting Eq. (\ref{eq:work out delta E}) into Eq. (\ref{eq:C_a_b}), we obtain the collision term, which consists of a 5-dimensional integration and may be calculated numerically by using the direct MC method on GPU.
With the help of the packages Ray and Numba, we can solve the Boltzmann equation (\ref{eq:boltzmann equation general form})
for all $2\leftrightarrow2$ scattering processes on the distributed
GPU clusters.

Before we present our numerical results, we would like to discuss a little more about the difference between integrating over $\mathbf{k}_{2}$ or $\mathbf{k}_{3}$ when we use the $\delta$-function.
The difference comes from solving $k_{1z}$ from the equation $E_{1}+E_{2}=E_{3}+E_{p}$.
In our first choice in Eq. (\ref{eq:work out delta E}), $k_{1z}$ is a function of $E_{3}+E_{p}$.
Since $E_{3}+E_{p}$ is always positive for all $\mathbf{k}_{3}$ and $\mathbf{p}$, the integration of the
collision term in Eq. (\ref{eq:boltzmann equation general form}) is stable.
If we integral out $\mathbf{k}_{3}$ in the $\delta$-function,
then $k_{1z}$ will become a function of $E_{2}-E_{p}$, which could flip its sign when we change $\mathbf{k}_{2}$ and $\mathbf{p}$.
Therefore, the integral in Eq. (\ref{eq:boltzmann equation general form})
using the second setup is not as stable as using the first setup.
We will discuss more on the stability of collision term in  Sec. \ref{subsec:Stability-test-of}.


\subsection{Particle number conservation and symmetrical sampling \label{subsec:Particle-number-conservation}}

\begin{figure}
\begin{centering}
\includegraphics[scale=0.5]{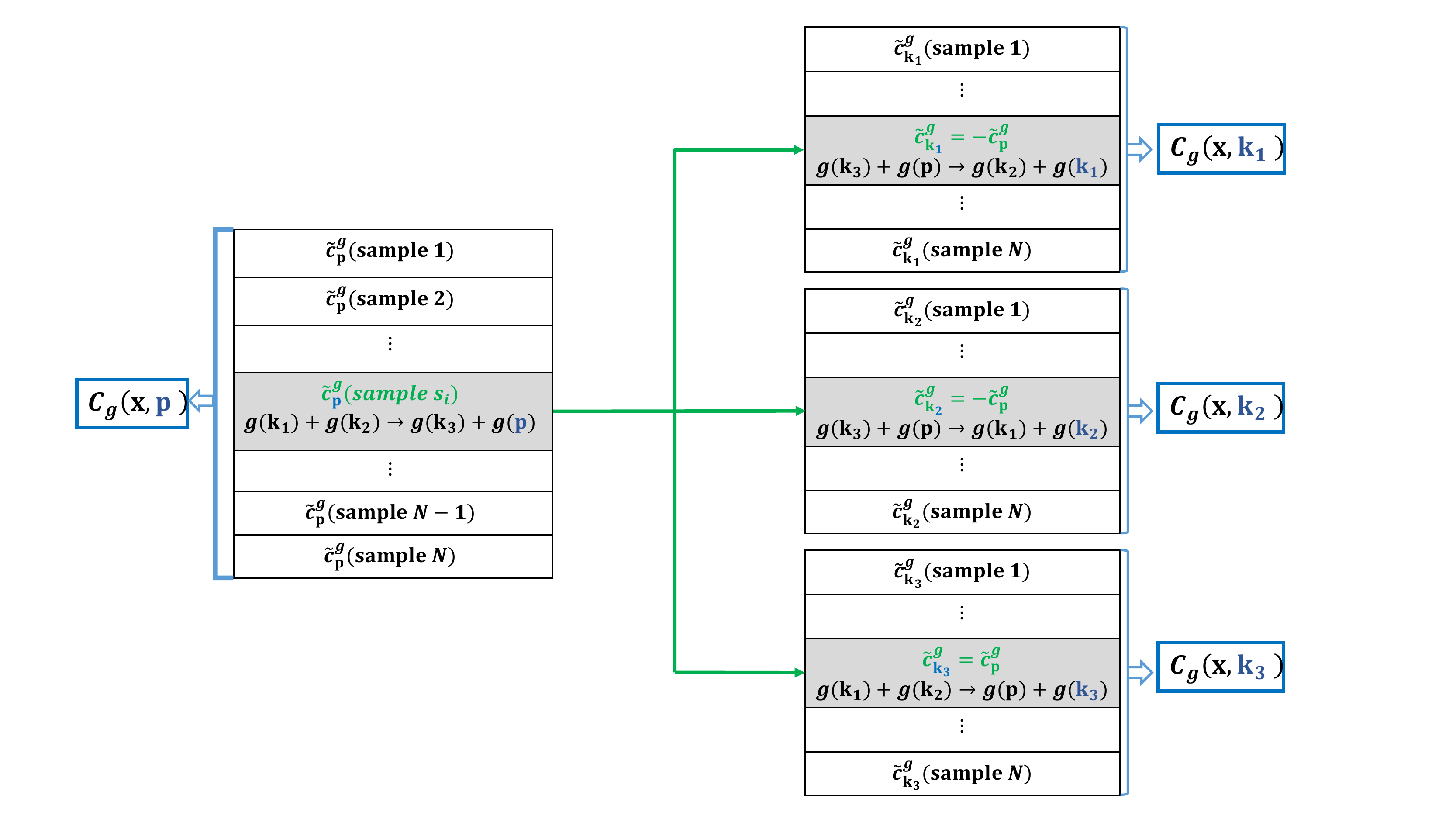}
\par\end{centering}
\caption{\label{fig:Procedure-of-symmetrical}Illustration of the ``symmetrical
sampling'' method. The sample $\tilde{c}_{\mathbf{p}}^{g}(\text{sample }s_{i})$
will be used by four integrations. This symmetrical reuse of samples
leads to a conserved particle number. We call this trick of using
sample $\tilde{c}_{\mathbf{p}}^{g}(\text{sample }s_{i})$ for all
four momentum grids as ``symmetrical sampling''. The sample points
have been quadrupled.}
\end{figure}

Since we use the direct MC method to compute the collision integral,
the total particle numbers are not strictly conserved in each time step due to the randomness
of MC sampling. Such non-conservation of particle numbers can accumulate with time and
may affect the result at later time steps. To ensure a strict particle
number conservation, we introduce a method named ``symmetrical sampling'' on GPUs.
Here we use the process of gluon scattering $g(\mathbf{k}_{1})+g(\mathbf{k}_{2})\rightarrow g(\mathbf{k}_{3})+g(\mathbf{p})$
as an example to illustrate the basic idea of our ``symmetrical sampling'' method. 
To calculate the collision term $C_{gg\rightarrow gg}(\mathbf{x},\mathbf{p})$
in Eq. (\ref{eq:C_a_b}), we need to sample a series values of the
integration variables $(k_{1x},k_{1y},k_{3x},k_{3y},k_{3z})$. The
collision term can be written as,
\begin{eqnarray}
C_{g(\mathbf{k}_{1})+g(\mathbf{k}_{2})\rightarrow g(\mathbf{k}_{3})+g(\mathbf{p})}(\mathbf{x},\mathbf{p}) & = & \int\tilde{c}_{\mathbf{p}}^{g}(k_{1x},k_{1y},k_{3x},k_{3y},k_{3z})dk_{1x}dk_{1y}dk_{3x}dk_{3y}dk_{3z}\nonumber \\
 & \simeq & \frac{V_{\text{domain}}}{N}\sum_{s_{i}=1}^{N}\tilde{c}_{\mathbf{p}}^{g}(\text{sample }s_{i}),\label{eq:MC C_gg}
\end{eqnarray}
where $V_{\text{domain}}$ is the volume of the integration domain, and the kernel $\tilde{c}_{\mathbf{p}}$ denotes,
\begin{eqnarray}
\tilde{c}_{\mathbf{p}}(k_{1x},k_{1y},k_{3x},k_{3y},k_{3z}) & = & \left[f_{k_{1}}^{a}f_{k_{2}}^{b}F_{k_{3}}^{c}F_{p}^{d}-F_{k_{1}}^{a}F_{k_{2}}^{b}f_{k_{3}}^{c}f_{p}^{d}\right]\times sym\nonumber \\
 &  & \times\sum_{i=1,2}\frac{1}{|Ja(k_{1z}^{i})|}\frac{1}{(2\pi)^{5}2E_{1}2E_{2}2E_{3}2E_{p}}|M_{ab\leftrightarrow cd}|^{2},\label{eq:ca}
\end{eqnarray}
where the symmetry factor $sym=1/2$ when the initial state is composed of two identical particles, and otherwise equals 1.
Given each of $(k_{1x},k_{1y},k_{3x},k_{3y},k_{3z})$ and $(p_{x},p_{y},p_{z})$, we can obtain the corresponding
$(k_{1z},k_{2x},k_{2y},k_{2z})$ and $\tilde{c}_{\mathbf{p}}(k_{1x},k_{1y},k_{3x},k_{3y},k_{3z})$.

Let us consider one specific sample $\tilde{c}_{\mathbf{p}}^{g}(\text{sample }s_{i})$
in Eq. (\ref{eq:MC C_gg}) (also see Fig. \ref{fig:Procedure-of-symmetrical}).
In the usual MC sampling, one will only add the contribution of $\tilde{c}_{\mathbf{p}}^{g}(\text{sample }s_{i})$
to $C_{\mathbf{k}_{1}+\mathbf{k}_{2}\rightarrow\mathbf{k}_{3}+\mathbf{p}}(\mathbf{x},\mathbf{p})$.
This sample will not influence the values of $C_{\mathbf{k}_{3}+\mathbf{p}\rightarrow\mathbf{k}_{1}+\mathbf{k}_{2}}(\mathbf{x},\mathbf{k}_{1})$,
$C_{\mathbf{k}_{3}+\mathbf{p}\rightarrow\mathbf{k}_{1}+\mathbf{k}_{2}}(\mathbf{x},\mathbf{k}_{2})$
and $C_{\mathbf{k}_{1}+\mathbf{k}_{2}\rightarrow\mathbf{k}_{3}+\mathbf{p}}(\mathbf{x},\mathbf{k}_{3})$,
for which one will compute their corresponding $\tilde{c}_{\mathbf{k}_{1}}^{g}$,
$\tilde{c}_{\mathbf{k}_{2}}^{g}$ and $\tilde{c}_{\mathbf{k}_{3}}^{g}$
separately.
Due to such independence of $C_{gg\rightarrow gg}(\mathbf{x},\mathbf{p})$ at each grid, one cannot achieve a strict particle number conservation in the MC approach.

To fix the issue of the particle number non-conservation, we reuse the value
of $\tilde{c}_{\mathbf{p}}^{g}(\text{sample }s_{i})$. Actually, for
a given set of $\mathbf{p},\mathbf{k}_{1}, \mathbf{k}_{2}, \mathbf{k}_{3}$
satisfying $\mathbf{k}_{1}+\mathbf{k}_{2}=\mathbf{k}_{3}+\mathbf{p}$,
the kernels $\tilde{c}_{\mathbf{p}}^{g}$, $\tilde{c}_{\mathbf{k}_{1}}^{g}$, $\tilde{c}_{\mathbf{k}_{2}}^{g}$
and $\tilde{c}_{\mathbf{k}_{3}}^{g}$ in different collisional integrals
are related to each other due to the symmetry in the scattering amplitude $|M_{ab\leftrightarrow cd}|^{2}$

\begin{equation}
\tilde{c}_{\mathbf{p}}^{g}=\tilde{c}_{\mathbf{k}_{3}}^{g}=-\tilde{c}_{\mathbf{k}_{2}}^{g}=-\tilde{c}_{\mathbf{k_{1}}}^{g},\qquad\textrm{for given \ensuremath{\mathbf{p}}, \ensuremath{\mathbf{k}_{1}}, \ensuremath{\mathbf{k}_{2}}, \ensuremath{\mathbf{k}_{3}} and \ensuremath{\mathbf{k}_{1}}+\ensuremath{\mathbf{k}_{2}}=\ensuremath{\mathbf{k}_{3}}+\ensuremath{\mathbf{p}} }.\label{eq:small c relation}
\end{equation}
The above relation can be easily seen from Eq. (\ref{eq:ca}).
If we switch particle 1 and $\mathbf{p}$ in Eq. (\ref{eq:ca}), we
only change the sign of the term $\left[f_{k_{1}}^{a}f_{k_{2}}^{b}F_{k_{3}}^{c}F_{p}^{d}-F_{k_{1}}^{a}F_{k_{2}}^{b}f_{k_{3}}^{c}f_{p}^{d}\right]$.
Therefore, in our program, we will also use the value of $\tilde{c}_{\mathbf{p}}^{g}(\text{sample }s_{i})$
for $C_{\mathbf{k}_{3}+\mathbf{p}\rightarrow\mathbf{k}_{1}+\mathbf{k}_{2}}(\mathbf{x},\mathbf{k}_{1})$,
$C_{\mathbf{k}_{3}+\mathbf{p}\rightarrow\mathbf{k}_{1}+\mathbf{k}_{2}}(\mathbf{x},\mathbf{k}_{2})$
and $C_{\mathbf{k}_{1}+\mathbf{k}_{2}\rightarrow\mathbf{k}_{3}+\mathbf{p}}(\mathbf{x},\mathbf{k}_{3})$
as well. We call this trick, to use the sample $\tilde{c}_{\mathbf{p}}^{g}(\text{sample }s_{i})$
for all four momentum grids, the ``symmetrical sampling'' method. This means that
the sample points have been quadrupled.

With our ``symmetrical sampling" trick, we can avoid the errors from the related samples in the collisional integrals at different momentum grids. Accordingly, we can obtain a strict particle number conservation.
In principle, one can apply this method in the direct MC sampling based on CPU approaches.
However, it usually takes lots of computing time and is hard to implement.
Fortunately, with the help of the feature in GPUs, named CUDA atomic operation \citep{cudaAutomicOperation,cudaAutomicOperation2},
the extra time for implementing symmetrical sampling is almost negligible.
As explained by the official documents \citep{cudaAutomicOperation3},
\emph{``Atomic operations are operations which are performed without interference from any other threads.
Atomic operations are often used to prevent race conditions, which are common problems in multithreaded applications.''}
In our case, each value of the array $\tilde{c}_{\mathbf{p}}^{g}$, whose element represents a specific $\tilde{c}_{\mathbf{p}}^{g}$
value at momenta $\mathbf{p}$, is saved in the global GPU memory. During the process of parallel evaluations,
at the same momenta $\mathbf{p}$, we obtain ``simultaneously'' many
values for $\tilde{c}_{\mathbf{p}}^{g}$ from different threads 
[we will also obtain the value of $\tilde{c}_{\mathbf{p}}^{g}$ from
$\tilde{c}_{\mathbf{k}_{3}}^{g}, -\tilde{c}_{\mathbf{k}_{2}}^{g}, -\tilde{c}_{\mathbf{k_{1}}}^{g}$ as discussed in Eq.(\ref{eq:small c relation})].
Since the accumulation of these values can only be performed sequentially, when one value of $\tilde{c}_{\mathbf{p}}^{g}$
in a GPU thread is calculated and being accumulated to this global memory array $\tilde{c}_{\mathbf{p}}^{g}$, all the other threads do not have the access of $\tilde{c}_{\mathbf{p}}^{g}$ at $\mathbf{p}$.
These processes in GPUs refer to CUDA ``atomic operation''.
Compared with parallel CPU manipulation, CUDA is extremely fast at performing this atomic operation, which enables a fast ``symmetrical sampling''.
Similar strategy has also been used in CPUs implementation to ensure particle number conservation, e.g. see Ref. \citep{Kurkela2014,Kurkela2015,Keegan2016,Kurkela2019,Kurkela2019a}.

We note that while the condition $\mathbf{k}_{1}+\mathbf{k}_{2}=\mathbf{k}_{3}+\mathbf{p}$ ensures that the integration always preserves energy and momentum conservation in all microscopic processes, the total energy as computed via Eq. (\ref{eq:energy momentum tensor}) might not be strictly conserved due to the discrete grids.
To explain the possible non-conservation of the total energy, we can take a close look at Eq. (\ref{eq:left handside}).
By integrating both sides of Eq. (\ref{eq:left handside}) over $d^{3}\mathbf{x}d^{3}\mathbf{p}/(2\pi)^{3}$
and using the definition of total particle number in Eq. (\ref{eq:particle numbers}),
we obtain the time variation of particle number
\begin{equation}
\frac{d}{dt}N=N_{a}\int d^{3}\mathbf{x}\frac{d^{3}\mathbf{p}}{(2\pi)^{3}}C[f],
\end{equation}
where $a=g,q(\bar{q})$. Our symmetrical sampling method in the collisional integral $C[f]$ guarantees the time reversal
symmetry in all microscopic processes, e.g. in Eq. (\ref{eq:small c relation}).
Such time reversal symmetry in collisional integral $C[f]$ makes the integral $\int\frac{d^{3}\mathbf{p}}{(2\pi)^{3}}C[f]$ exactly
zero numerically \citep{DeGroot:1980dk,Pu:2011vr}, which ensures the strict particle number conservation.

Similarly, by integrating Eq. (\ref{eq:left handside}) over $d^{3}\mathbf{x}d^{3}\mathbf{p}/(2\pi)^{3}$ with the multiplication of $p^{0}$ on both sides and using the definition of energy-momentum tensor in Eq. (\ref{eq:energy momentum tensor}),
we get the time variation of total energy,
\begin{equation}
\frac{d}{dt}T^{00}=N_{a}\int d^{3}\mathbf{x}\frac{d^{3}\mathbf{p}}{(2\pi)^{3}}p^{0}C[f]+\int d^{3}\mathbf{x}S_{ex}^{0},
\end{equation}
where $S_{ex}^{\mu}$ is the source term in Eq. (\ref{eq:current conservation}).
For simplicity, let us assume the mass is constant, then the source
term vanishes. Although the time reversal symmetry still holds, the
errors could come from the discrete grids for $p^{0}$. Therefore,
eventually, the integral $\int\frac{d^{3}\mathbf{p}}{(2\pi)^{3}}p^{0}C[f]$
is not strictly zero numerically. On the other hand, from the above analysis,
we could expect that the errors for the non-conservation of total energy will
decrease if we increase the number of grids.
We will address this point in more details in Sec. \ref{subsec:Total-energy-conservation}.

\section{Test of program and time evolution in coordinate and momentum space
\label{sec:performance}}

In this section, we will first test several aspects of our program, and then show the time evolution of the distribution in both coordinate and momentum spaces.
The stability of the collision integrals will be tested in Sec. \ref{subsec:Stability-test-of}.
The check of particle number conservation will be presented in Sec. \ref{subsec:Test-of-Particle}, and our results show that the particle number is strictly conserved.
The total energy conservation is checked in Sec. \ref{subsec:Total-energy-conservation}.
It is found that even though the total energy is not strictly conserved, the numerical errors will decrease very fast with increasing the number of the grid.
Finally, we will present the time evolution of the systems in both coordinate and momentum spaces for pure gluons, pure quarks, and the mixture of quarks and gluons in Sec. \ref{subsec:Evolution-of-coordinates} and \ref{subsec:Evolution-in-momentum}.
As is expected, the system tends to be homogenous in coordinate space and become thermalized in momentum space.
We also find indications of gluon condensation in the soft region.

\subsection{Test of the stability of the collision integrals \label{subsec:Stability-test-of}}

\begin{figure}
\begin{centering}
\includegraphics[scale=0.45]{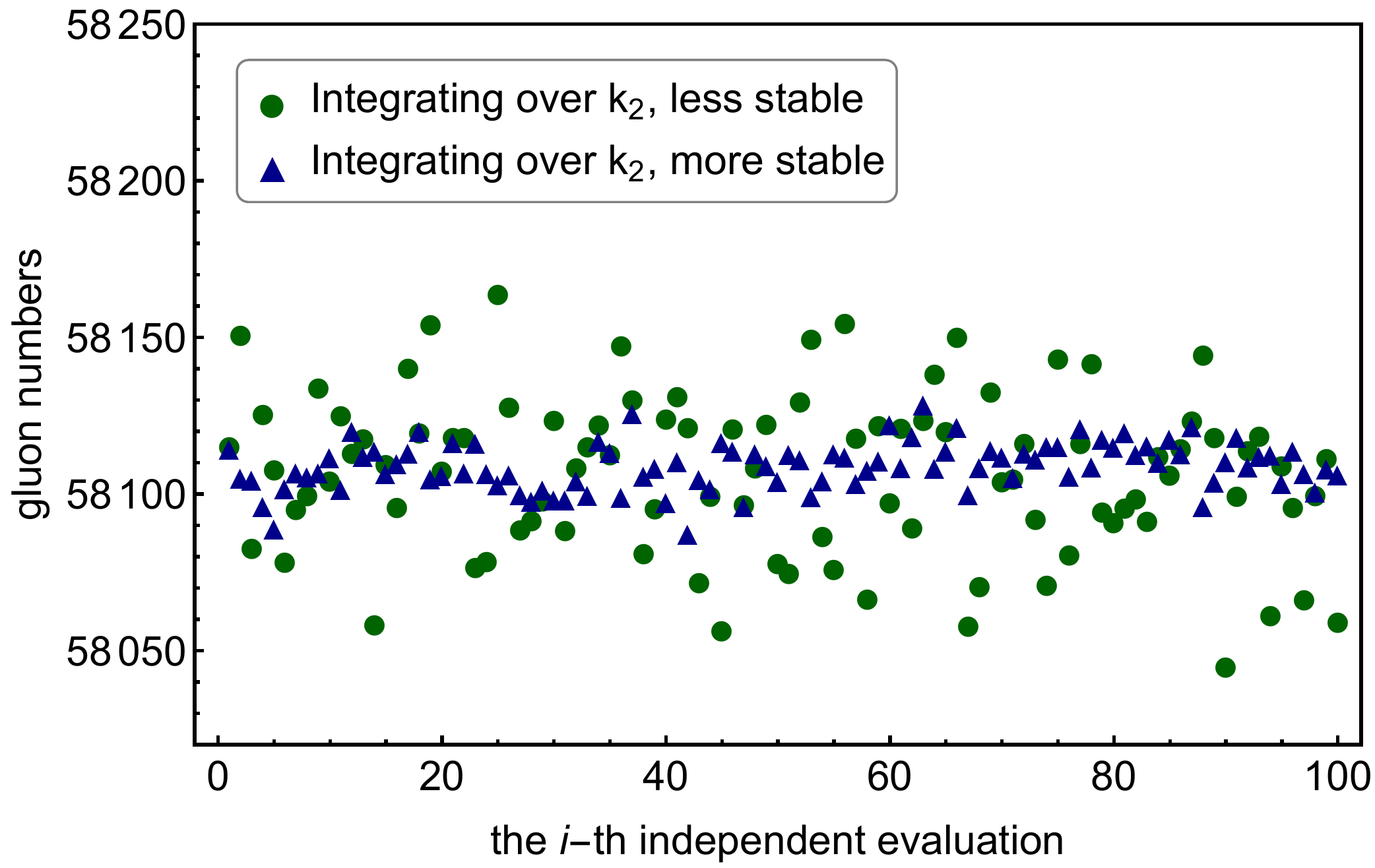}
\par\end{centering}
\caption{\label{fig:Comparison-of-the} Test for the stability of two approaches dealing with the delta function.
Each data point represents the gluon numbers for the distribution function $f_{g}(t_{0}+dt,\mathbf{x},\mathbf{p})$,
where $t_{0}=0\text{fm}$ and $dt=0.01\text{fm}$.
The horizontal axis denotes the $i$-th independent evaluation of the evolution from $t_{0}$ to $t_{0}+dt$. The initial distribution function $f_{g}(t_{0},\mathbf{x},\mathbf{p})$, drawn from random values between {[}0,1{]}, is kept the same for all data points in the figure.
The phase space box is of size $[-3\text{fm},3\text{fm}]^{3}\times[-2\text{GeV},2\text{GeV}]^{3}$.
We take 40 grids for each momentum freedom, the spatial grid is set to 1, and $\alpha_{s}$ = 0.3.
For each integration point $C_{g}(\mathbf{x},\mathbf{p})$, we have sampled 100 points to perform a direct MC integration.
The matrix element for gluon-gluon scattering is in Tab. \ref{tab:Matrix-elements-squared}.
The blue triangle and green circle points stand for the results of integrating over $\mathbf{k}_{2}$ and $\mathbf{k}_{3}$ in Eq. (\ref{eq:delta function}), respectively.}
\end{figure}

As mentioned in Sec. \ref{sec:numerical-method-involved}, there are two ways to integrate over the $\delta$-function in the collision term. Different approaches of handling the $\delta$-function can affect the numerical stability of the collision term.
Here we use gluon-gluon scatterings to illustrate such difference.

In Fig. \ref{fig:Comparison-of-the}, we show the results from two approaches
using the same initial distribution for gluons. Each data point
represents the numbers of the gluons associated with the distribution function $f_{g}(t_{0}+dt,\mathbf{x},\mathbf{p})$
where $t_{0}=0\text{fm}$ and $dt=0.01\text{fm}$.
The blue triangle points, labelled as ``more stable'', stand for particle numbers obtained by using $\mathbf{k}_{2}=\mathbf{k}_{3}+\mathbf{p}-\mathbf{k}_{1}$.
The green circle points, labelled as ``less stable'', denote the results by using $\mathbf{k}_{3}=\mathbf{k}_{1}+\mathbf{k}_{2}-\mathbf{p}$.
We can see that the green circle points spread over a relatively larger area than the blue triangle ones,
which means that the errors in the ``less stable'' method are relatively larger than the ``more stable'' ones.
Therefore, using $\mathbf{k}_2=\mathbf{k}_{3}+\mathbf{p}-\mathbf{k}_{1}$ to integrate over $d\mathbf{k}_{2}$ is a more stable method, consistent with our previous argument in Sec. \ref{sec:numerical-method-involved}.

\subsection{Test of particle number conservation \label{subsec:Test-of-Particle}}

\begin{figure}
\begin{centering}
\includegraphics[scale=0.39]{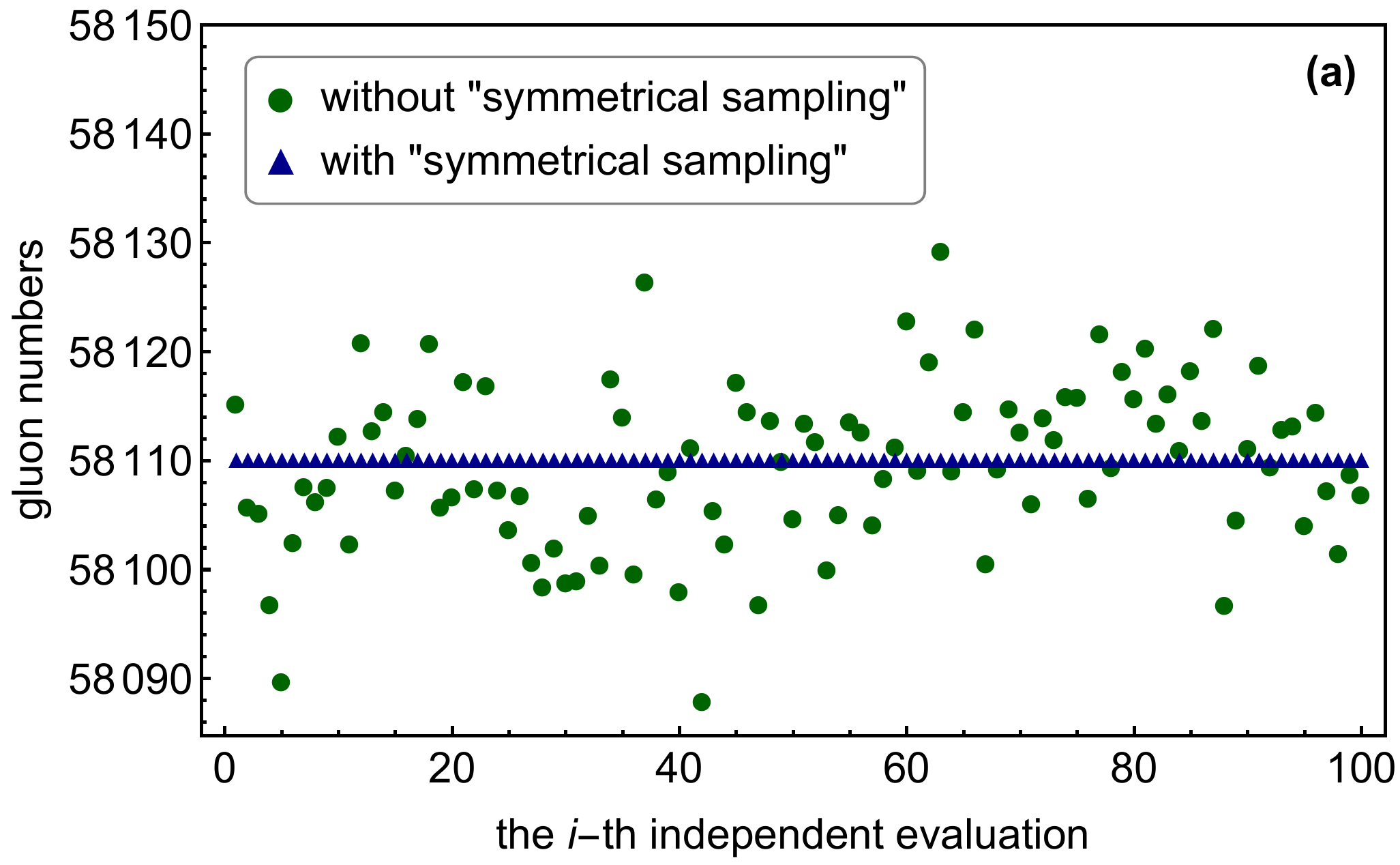}\includegraphics[scale=0.39]{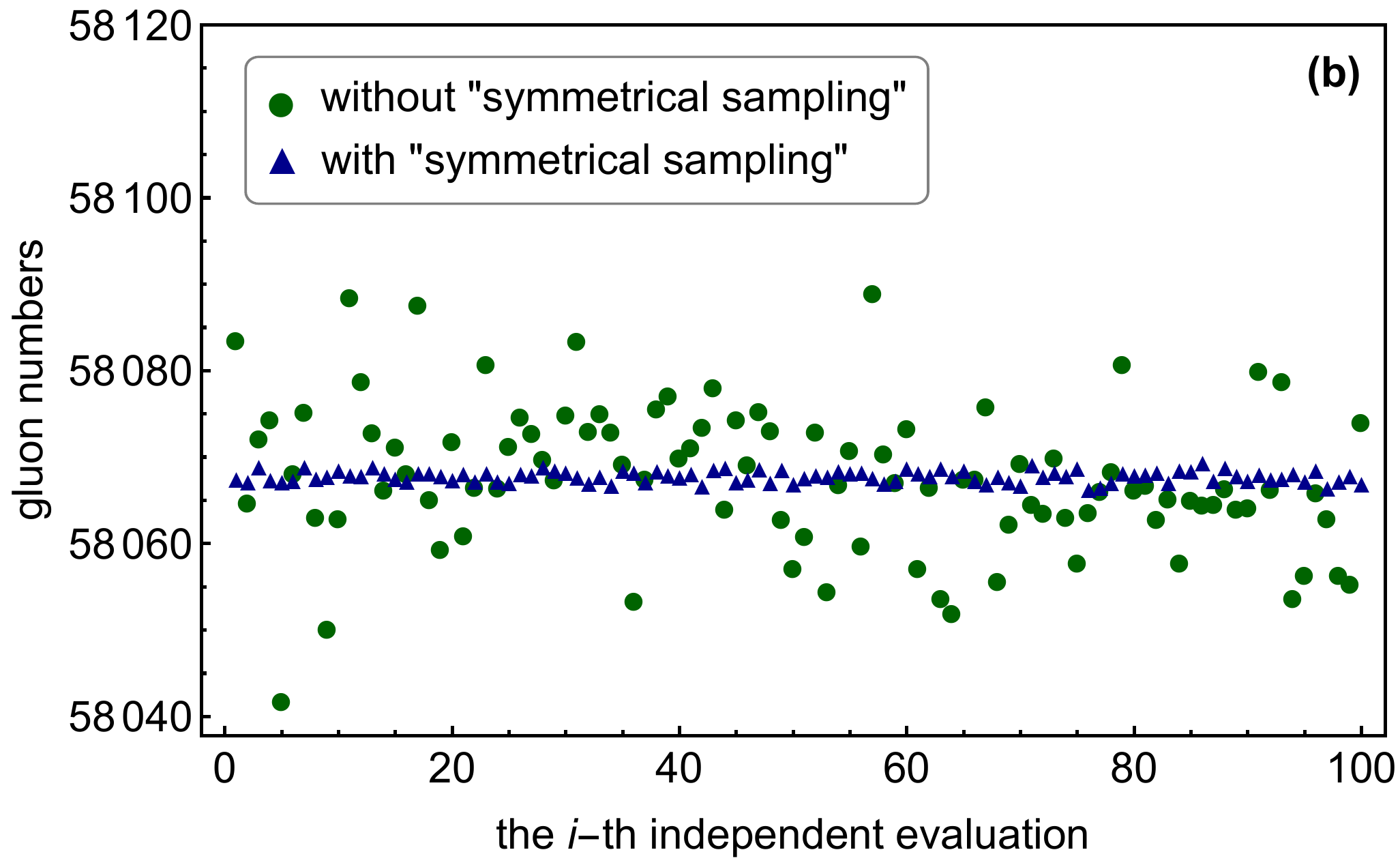}
\par\end{centering}
\begin{centering}
\includegraphics[scale=0.39]{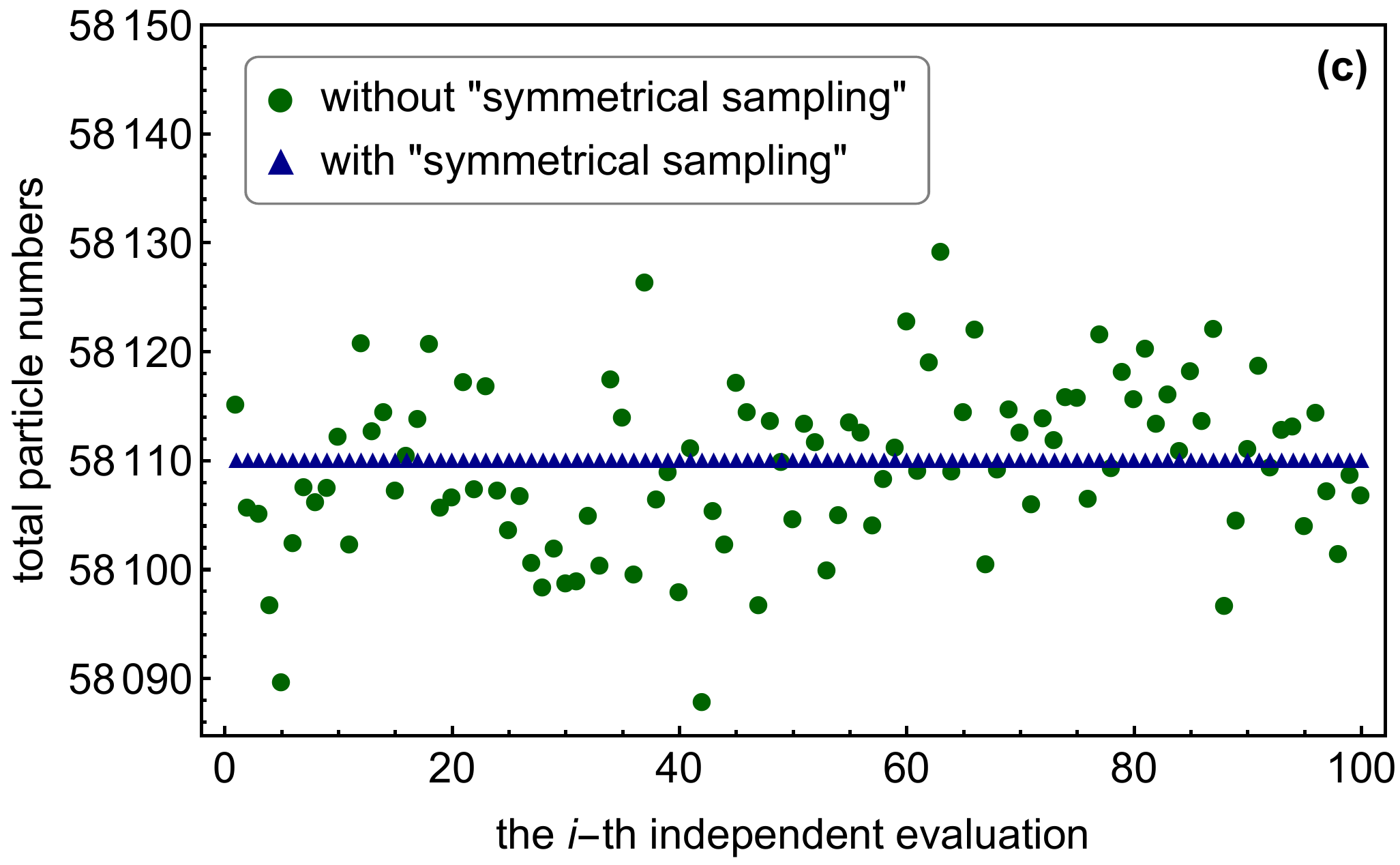}$\ \ \ $\includegraphics[scale=0.39]{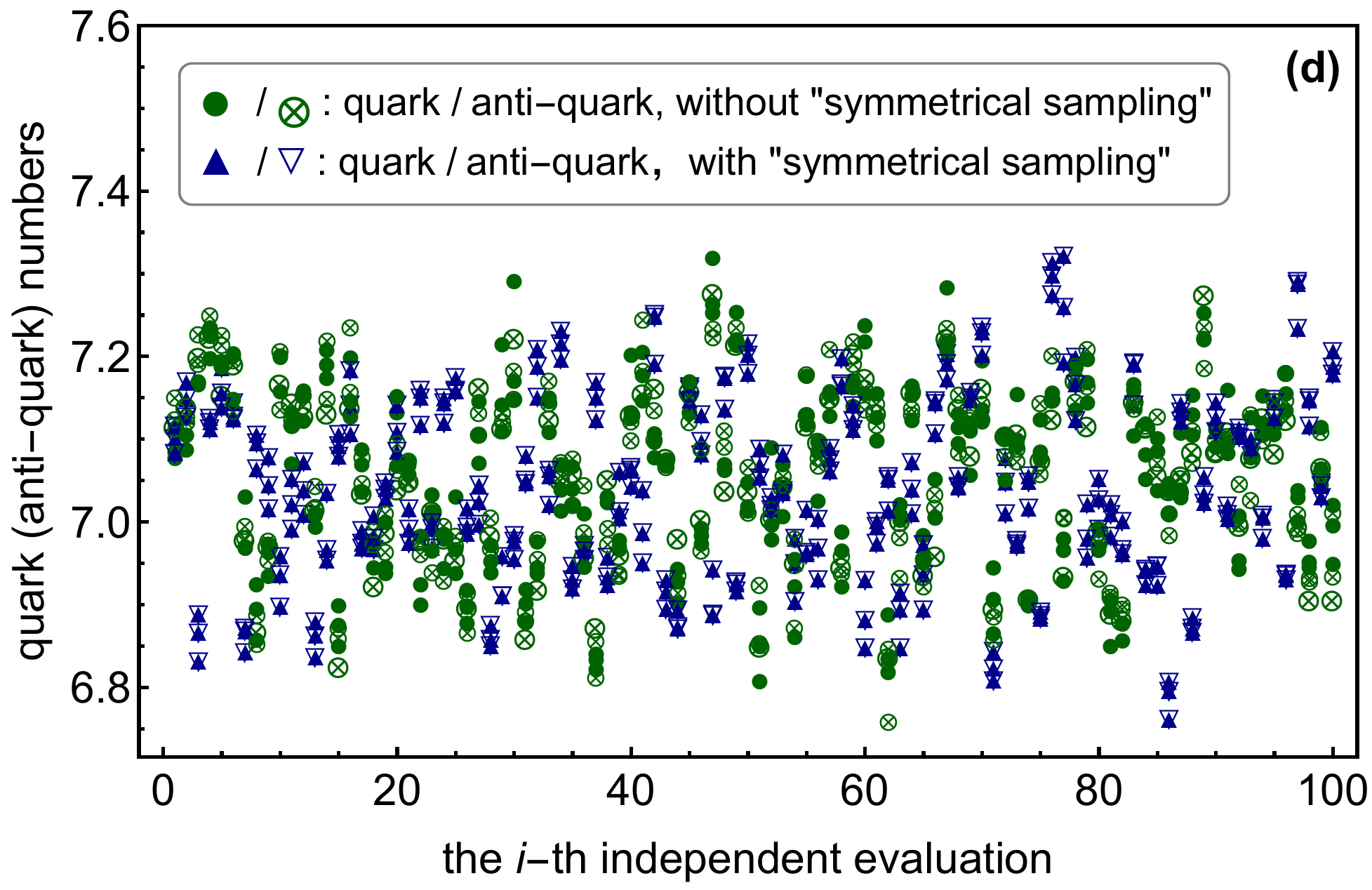}
\par\end{centering}
\caption{\label{fig:Comparison-between-with} Number of particles for the cases with and without ``symmetrical sampling''.
Each data point represents the particle numbers for the distribution function $f_{a}(t_{0}+dt,\mathbf{x},\mathbf{p})$ using the same initial configuration as in Fig. \ref{fig:Comparison-of-the}.
The horizontal axis denotes the $i$-th independent evaluation of the evolution from $t_{0}$ to $t_{0}+dt$.
The blue triangle and green circle points denote the results with and without symmetrical sampling, respectively.
The panel (a) is for the pure gluon case and the panels (b), (c) and (d) are for the cases including the scatterings of quarks and gluons.}
\end{figure}

In Fig. \ref{fig:Comparison-between-with}, we show the particle numbers for different parton species.
For simplicity, we label the results obtained by our symmetrical sampling method in Sec. \ref{subsec:Particle-number-conservation}
as ``with symmetrical sampling'', while the results from direct MC simulations are labelled as ``without symmetrical sampling''.
In the figures, the blue triangle and green circle points stand for the cases with and without symmetrical sampling, respectively.
In the upper panel, we show the gluon numbers for cases of pure gluon scattering and quark-gluon scattering in Fig. \ref{fig:Comparison-between-with} (a) and Fig. \ref{fig:Comparison-between-with} (b), respectively.
We can see that for pure gluon case, the gluon number is conserved when using our symmetrical sampling method.
In Fig. \ref{fig:Comparison-between-with} (b), since the gluons can be converted to quarks and anti-quarks, there are some variations of the gluon numbers.
Figure \ref{fig:Comparison-between-with} (d) shows that there are also variations for the numbers of quarks and anti-quarks.
However, as shown in Fig. \ref{fig:Comparison-between-with} (c), the total number of particles, including quarks, anti-quarks and gluons is conserved for $2\rightarrow2$ scatterings when using our symmetrical sampling method.

\begin{figure}
\begin{centering}
\includegraphics[scale=0.39]{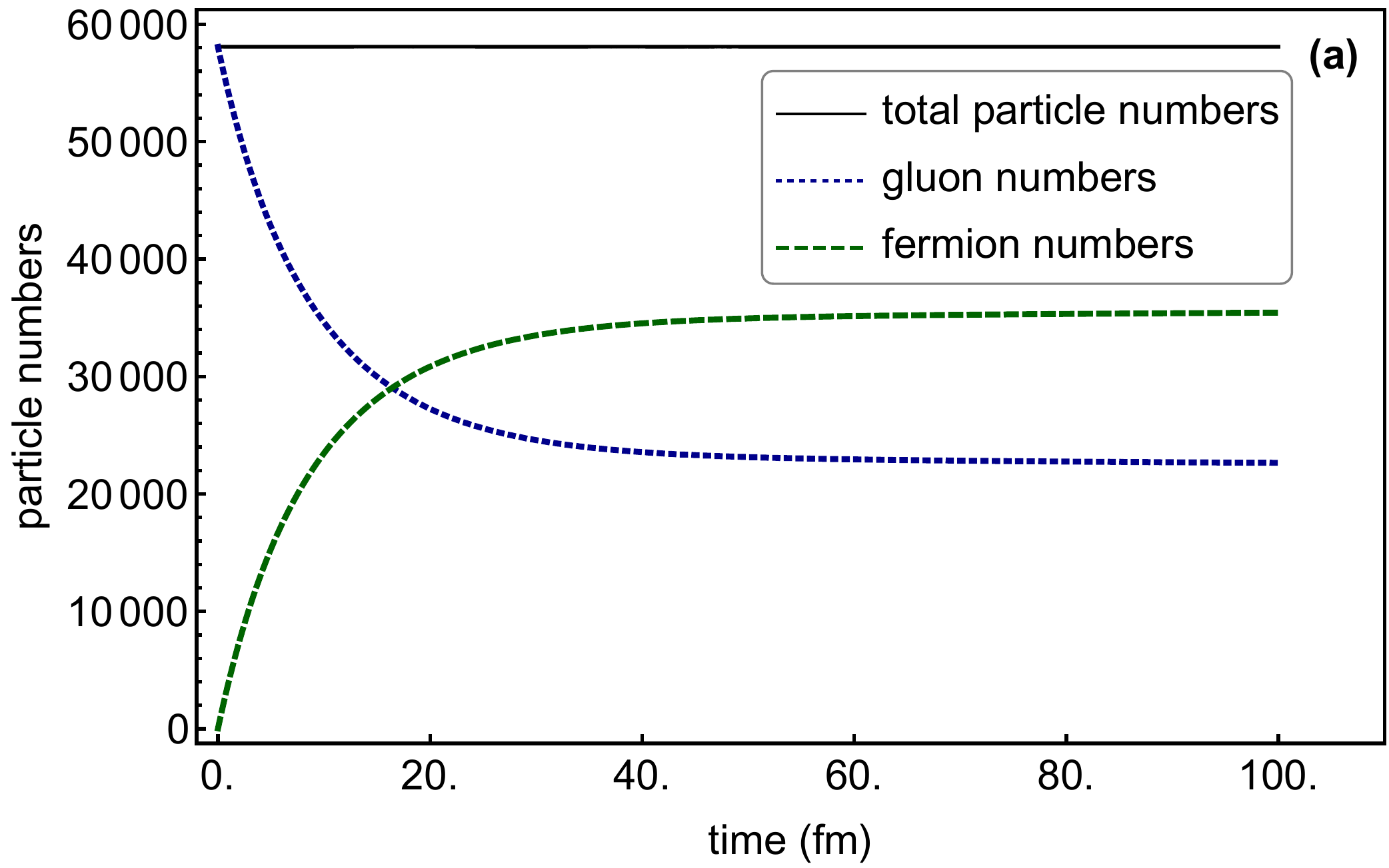}\includegraphics[scale=0.39]{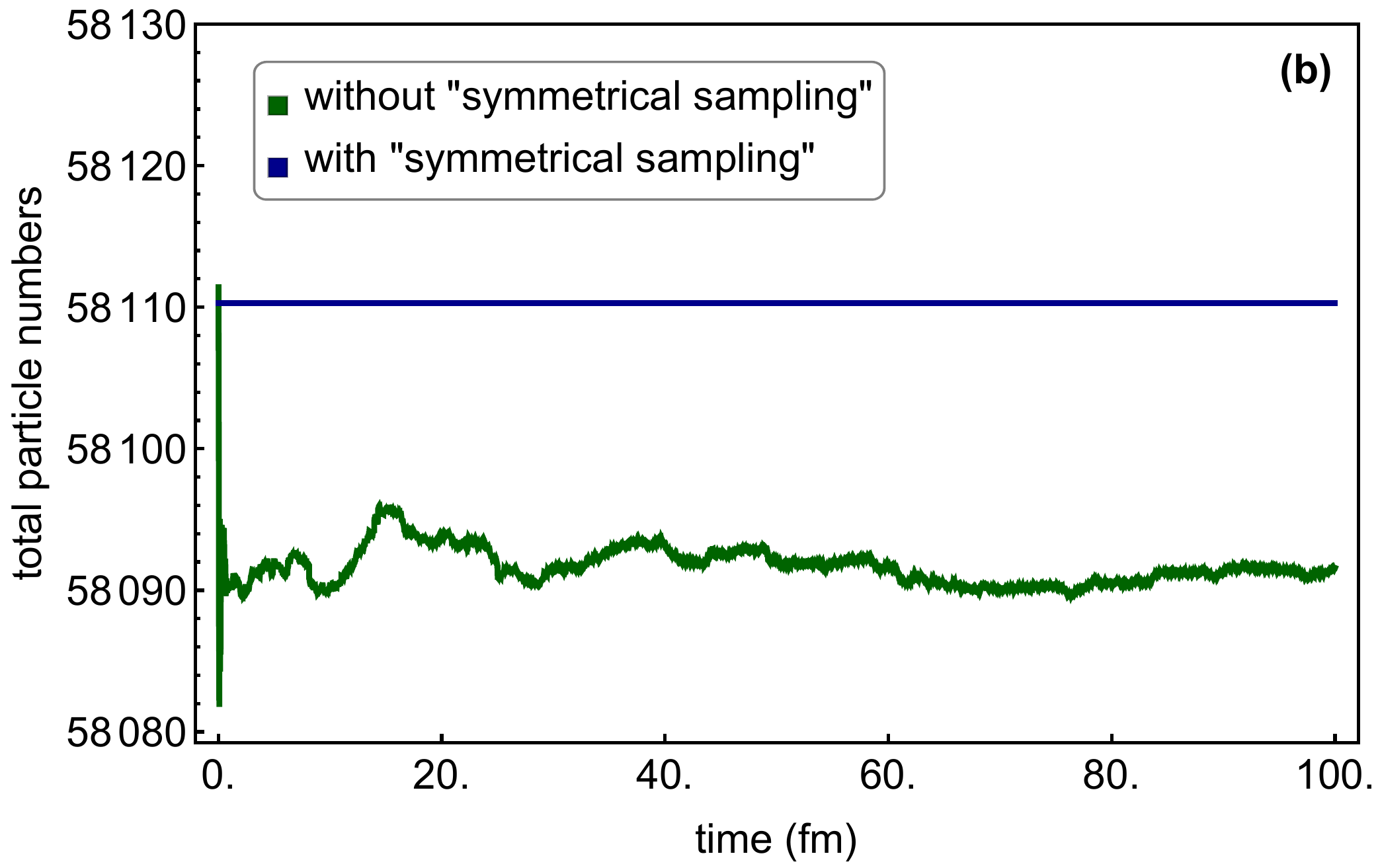}
\par\end{centering}
\caption{\label{fig:Comparison-between-with-2}
Number of particles of different species as a function of evolution time for the cases with and without ``symmetrical sampling''.
The parameters for $f_{g/q(\bar{q})}(t,\mathbf{x},\mathbf{p})$ are the same as in Fig. \ref{fig:Comparison-of-the} and \ref{fig:Comparison-between-with}.
Panel (a) plots the particle numbers for both fermions and gluons using the symmetrical sampling.
Panel (b) compares the total particle number as a function of evolution time with and without using the symmetrical sampling, as shown by the blue and green lines, respectively.
On one Nvidia Tesla V100 card, the entire evaluations take 6456 and 8198 seconds for the cases without and with the symmetrical sampling.}
\end{figure}

In Fig. \ref{fig:Comparison-between-with-2}, we show the numbers of gluons and quarks (anti-quarks) as a function of the evolution time. Since we initialize the system with all gluons, we find that gluons tend to convert into quarks and anti-quarks during the evolution, see in Fig. \ref{fig:Comparison-between-with-2} (a).
After some time, both gluon and quark numbers tend to achieve equilibrium.
While the individual parton numbers are changing with time, the total particle number is strictly conserved during the evolution when our symmetrical sampling method is employed. 
Without the symmetrical sampling, the conservation of the total particle number can be violated by a small amount (about 0.03\%) as shown in Fig. \ref{fig:Comparison-between-with-2} (b).

\subsection{Test of energy conservation \label{subsec:Total-energy-conservation}}

\begin{figure}
\begin{centering}
\includegraphics[scale=0.4]{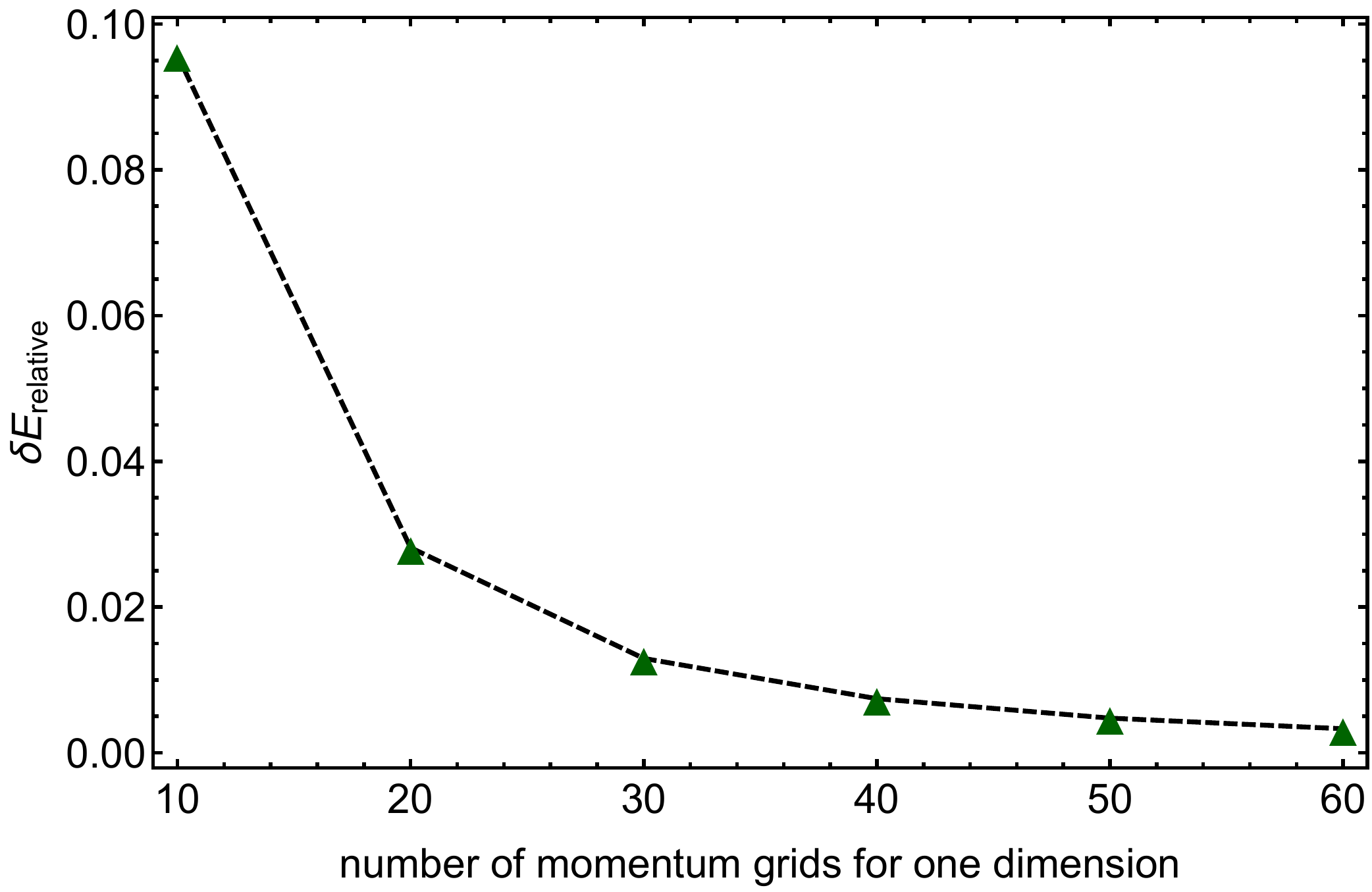}
\par\end{centering}
\caption{\label{fig:Total-energy-deviation}\textcolor{red}{{} }
The deviation of the total energy as a function of the number of momentum grids,  for a pure gluon system.
Here, the number for spatial grid is taken to be one, $\alpha_{s}$ = 0.3, the phase space box is of size $[-3\text{fm},3\text{fm}]^{3}\times[-2\text{GeV},2\text{GeV}]^{3}$ and the time step $dt=0.0001\text{fm}$.
The initial gluon distribution is chosen as a step function, $f_{g,\textrm{intial}}(\mathbf{p})=0.5\times\theta(1-|\mathbf{p}|/Q_{s})|$
with $Q_{s}=1.5\text{GeV}$ and $m_{g}^{2}=0.5\text{GeV}$.}
\end{figure}

In implementing the MC integration of the collision term, we can use the symmetrical sampling method to ensure the strict particle number conservation. However, the conservation of total particle number does not guarantee the strict conservation of total energy numerically due to the discrete grids in the numerical calculation.
When we calculate the total energy with Eq. (\ref{eq:energy momentum tensor}), the smooth $p^{0}$ is approximated by  discrete grids.  This discretization will affect the numerical evaluation of the total energy.
Such effect can be seen in Fig. \ref{fig:Total-energy-deviation}, where a constant gluon mass $m_{g}^{2}=0.5\text{GeV}^{2}$ is used.
To quantify the violation of energy conservation, here we introduce the following quantity $\delta E_{\text{relative}}$,
\begin{equation}
\delta E_{\text{relative}}=\frac{|E_{t=1\text{fm}}-E_{t=0\text{fm}}|}{E_{t=0\text{fm}}}.
\end{equation}
In Fig. \ref{fig:Total-energy-deviation}, we plot $\delta E_{\text{relative}}$ as a function of the number of momentum grids.
In this test, we choose the initial gluon distribution as a step function, $f_{g,\textrm{intial}}(\mathbf{p})=0.5\times\theta(1-|\mathbf{p}|/Q_{s})|$ with $Q_{s}=1.5\text{GeV}$.
We find that the deviation $\delta E_{\text{relative}}$ decreases fast as a function of the number of the grids.
From the figure, we can see that when the number of grids in momentum space is taken as $30$, $\delta E_{\text{relative}} \sim 0.015$, i.e. the fluctuation of total energy computated from Eq. (\ref{eq:energy momentum tensor}) is about $1.5\%$ at $t=1\textrm{fm/c}$.
Such small deviation is mainly caused by the discretization of momentum.
Of course, the total energy is still conserved physically for this case since we have used the condition $\mathbf{k}_{1}+\mathbf{k}_{2}=\mathbf{k}_{3}+\mathbf{p}$ in the computation of the collision term.

\begin{figure}
\begin{centering}
\includegraphics[scale=0.42]{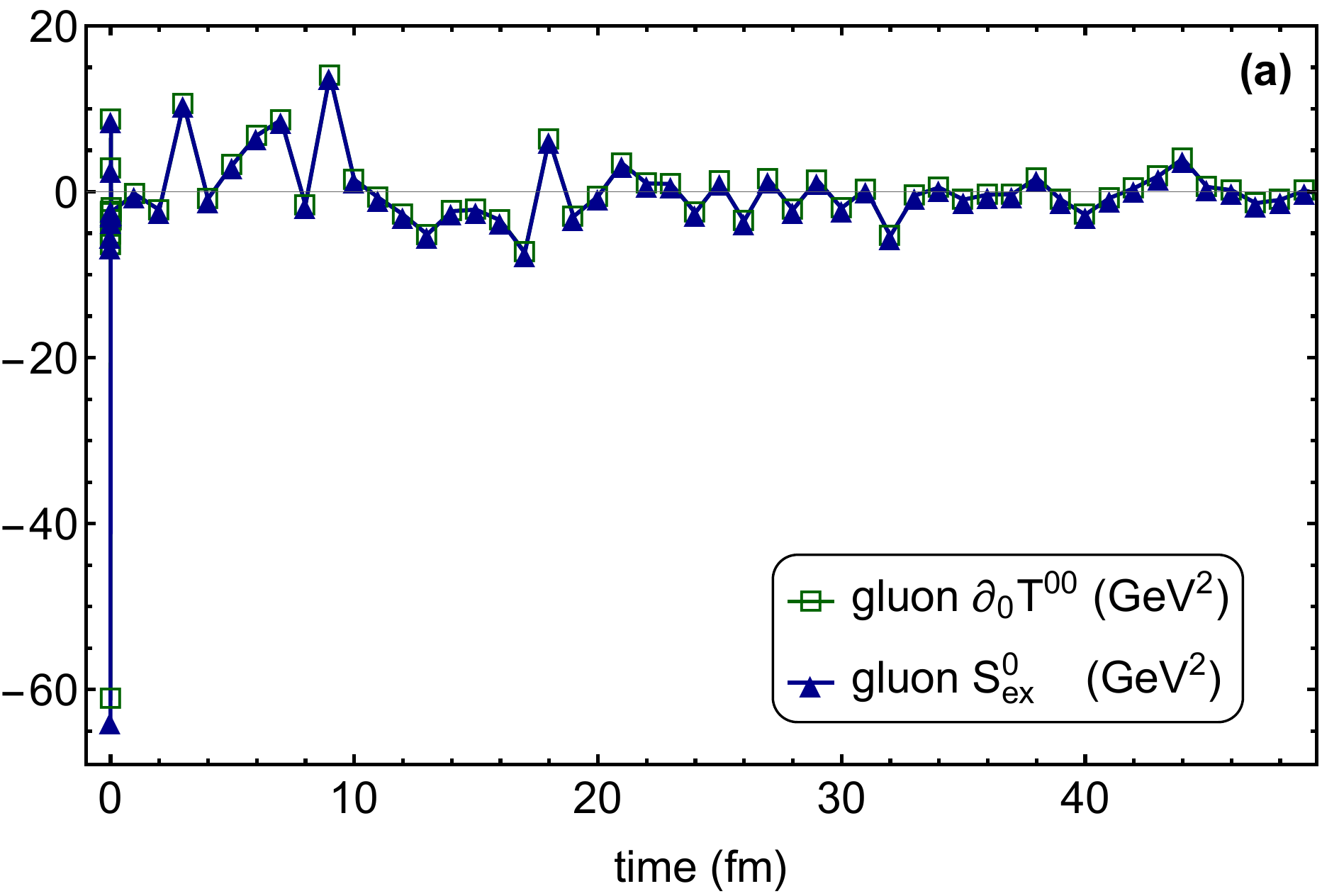}\includegraphics[scale=0.42]{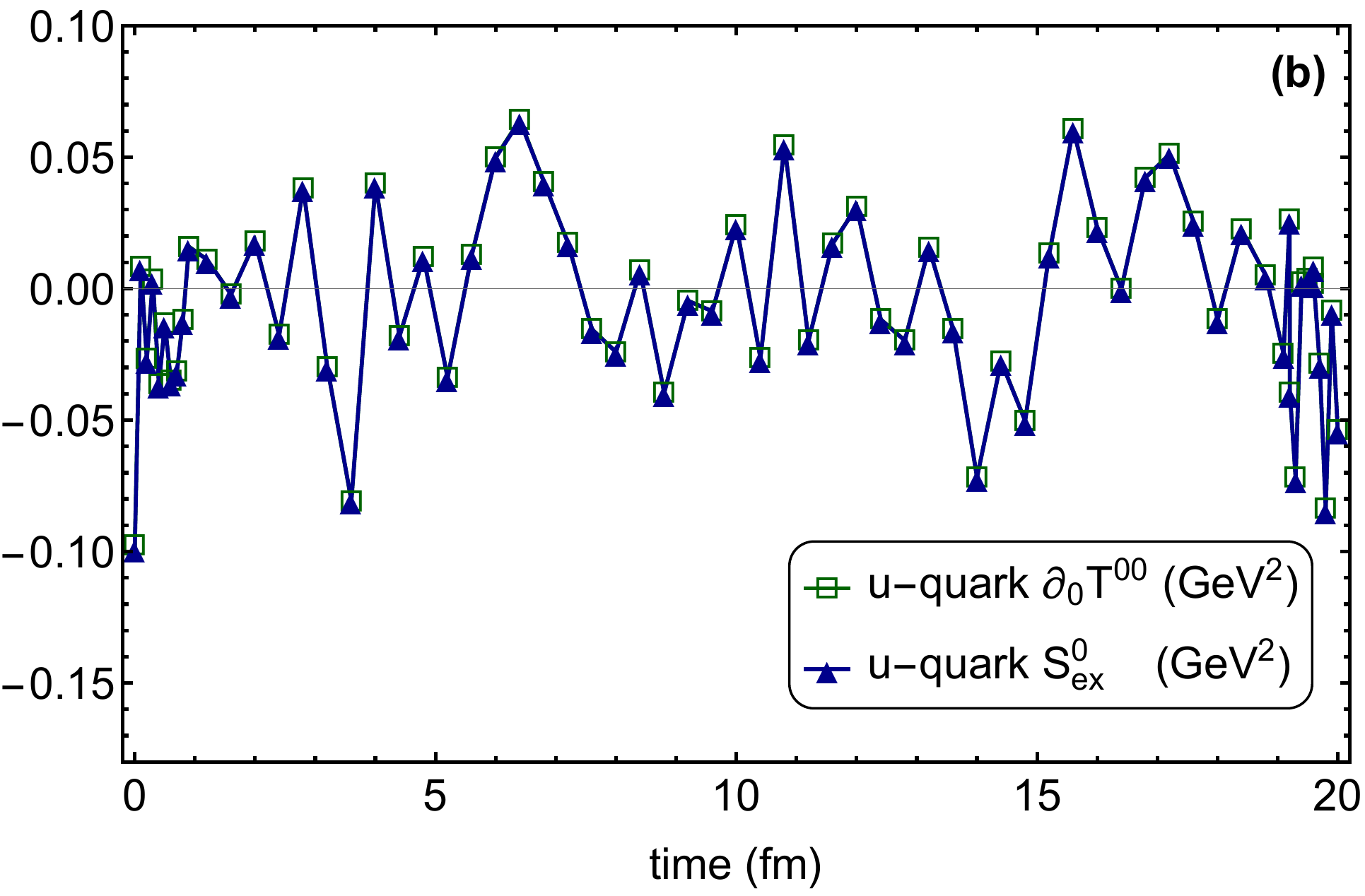}
\par\end{centering}
\caption{\label{fig:Check-.-The}
Time variation of the kinetic energy $\partial_{0}T_{kin}^{00}$ and the zero component of source term $S_{ex}^{0}$, as a function of evolution time.
Panels (a) and (b) show the results for pure gluon and quark systems, respectively.
The initial distribution function is given by $f_{g/u,\textrm{intial}}(\mathbf{p})=0.5\times\theta(1-|\mathbf{p}|/Q_{s})|$ with $Q_{s}=1.5\textrm{GeV}.$
We set $\alpha_{s}=0.3$, $dt=0.00005\textrm{fm}$ for gluons and $dt=0.001\textrm{fm}$ for quarks.}
\end{figure}

Now we consider the case of dynamical mass as computed by Eq. (\ref{eq:mg},
\ref{eq:mq}). For simplicity, we focus on the systems of pure gluons or $u$ quarks that are homogenous in coordinate space.
Then Eq. (\ref{eq:current conservation}) reduces to,
\begin{eqnarray}
\partial_{0}T_{kin}^{00} & = & S_{ex}^{0}=\frac{1}{2}\partial^{0}m_{a}^{2}\int\frac{d^{3}\mathbf{p}}{(2\pi)^{3}E_{p}^{a}}N_{a}f_{p}^{a}.\label{eq:check energy conservation}
\end{eqnarray}
Here the term $\partial_{0}T_{kin}^{00}(x)$ is evaluated directly via $[T^{00}(t+dt)-T^{00}(t-dt)]/(2dt)$.
In Fig. \ref{fig:Check-.-The}, we show the time variation of the kinetic energy $\partial_{0}T_{kin}^{00}$ and the zero component
of source term $S_{ex}^{0}$.
We can see that except for the first few steps, two terms ($\partial_{0}T_{kin}^{00}$ and $S_{ex}^{0}$) are almost identical.
For gluons as shown by \ref{fig:Check-.-The} (a), the value of $\partial_{0}T_{kin}^{00}(x)$ oscillates drastically for the first few time steps, but after $t=20\text{fm}$, it slightly fluctuates around the zero value, which indicates the reach of nearly thermal equilibration.
This result demonstrates that the change of the total energy comes from the source term.

The above consistency check in Fig. \ref{fig:Check-.-The} means that our results satisfy Eq. (\ref{eq:check energy conservation}) automatically. But this does not mean that the total energy is physically conserved since the masses are dynamically changing with space and time.
In principle, for a single flavor case, one can rewrite the right-hand side of Eq. (\ref{eq:current conservation}) as a total derivative term, then define a modified conserved total energy-momentum tensor as follows \citep{Jeon1995},
\begin{equation}
T_{total}^{\mu\nu}=T_{kin}^{\mu\nu}-\frac{1}{4}m_{a}^{2}\int\frac{d^{3}\mathbf{p}}{(2\pi)^{3}E_{p}^{a}}N_{a}f_{p}^{a},
\end{equation}
where the definition of squared mass in Eq. (\ref{eq:mg}, \ref{eq:mq}) is used to derive the second term.
For a multiple flavor case, one usually cannot get a modified conserved energy-momentum tensor.

For the case of dynamical mass, there may be another source of numerical error originating from the definition of mass in Eq. (\ref{eq:mg}, \ref{eq:mq}), apart from the discrete grids.
To obtain the squared mass, we need to integrate over $\mathbf{p}$ with $E_{p}=\sqrt{\mathbf{p}^{2}+m^{2}}$, which depends on the dynamical mass.
At the first step of the time evolution, we use $\mathbf{|}\mathbf{p}|$ instead of $E_{p}$ to derive the dynamic mass.
This tiny difference may also be a source of numerical errors.
Despite the above mentioned numerical errors, the conservation of the total energy computed via Eq. (\ref{eq:energy momentum tensor}) can be archived up to 99.8\% in Fig. \ref{fig:Check-.-The}.

\subsection{Evolution of distribution functions and dynamical masses in coordinate space
\label{subsec:Evolution-of-coordinates}}

\begin{figure}
\begin{centering}
\includegraphics[scale=0.28]{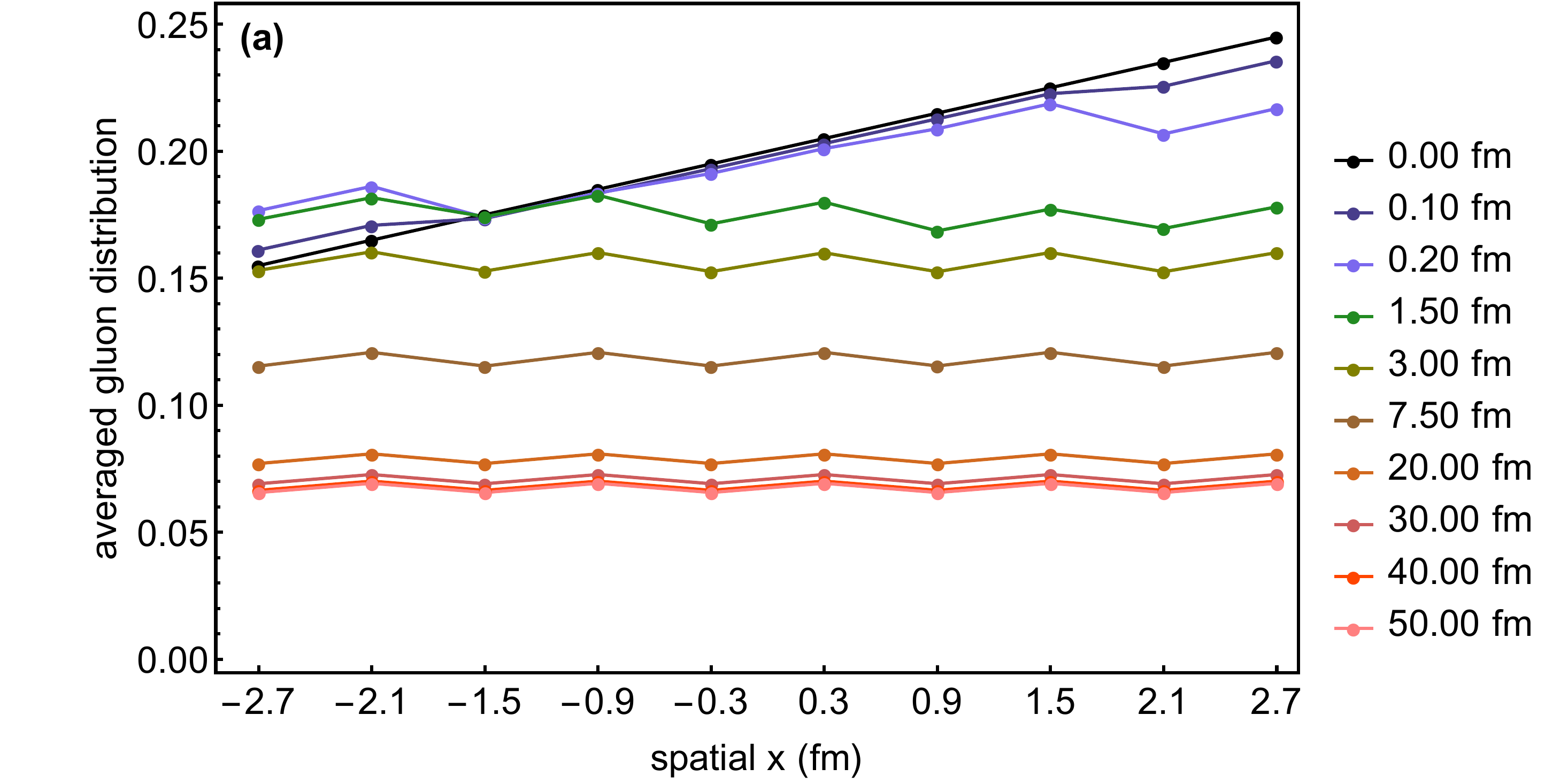}\includegraphics[scale=0.28]{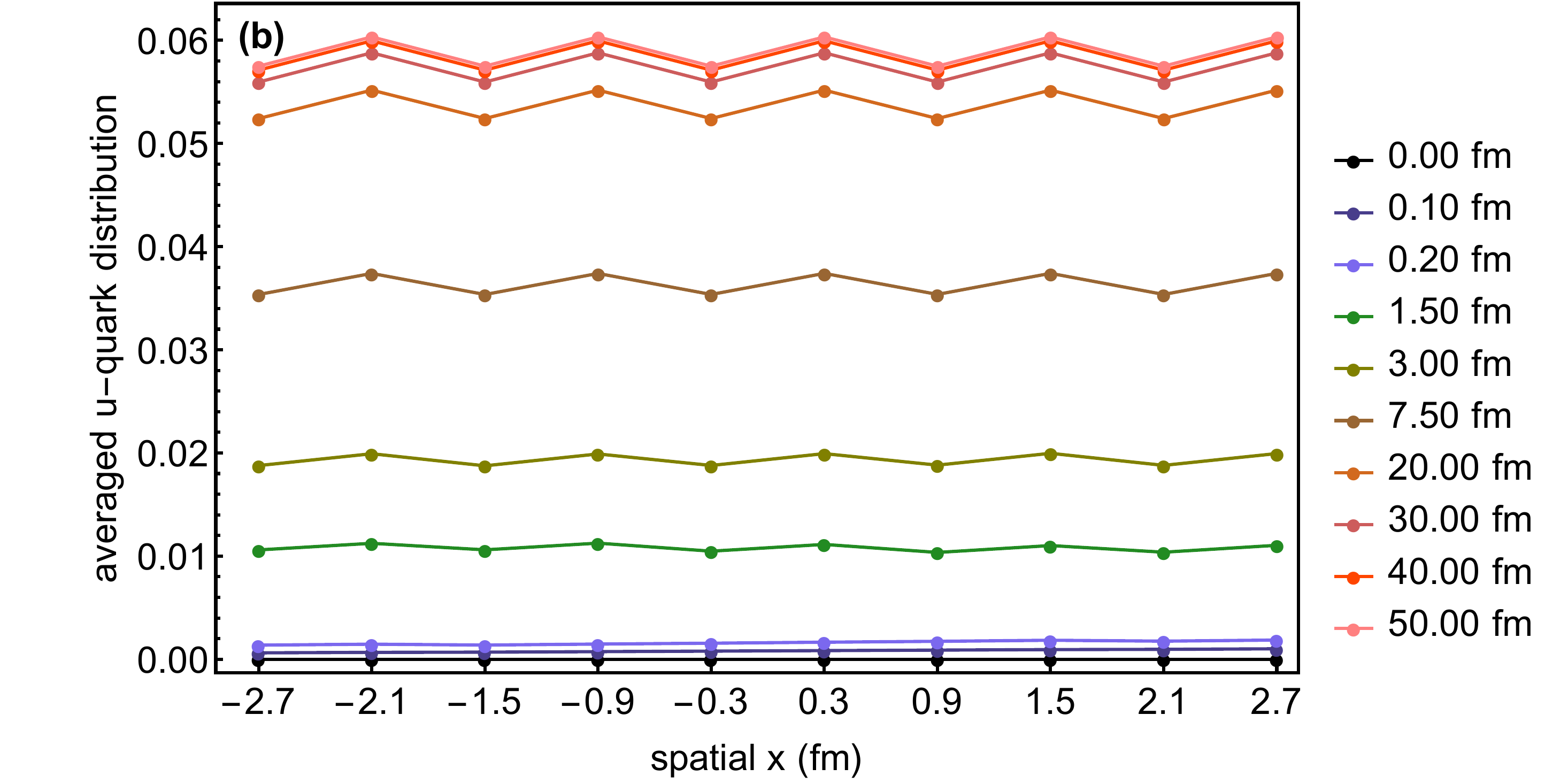}
\par\end{centering}
\begin{centering}
\includegraphics[scale=0.28]{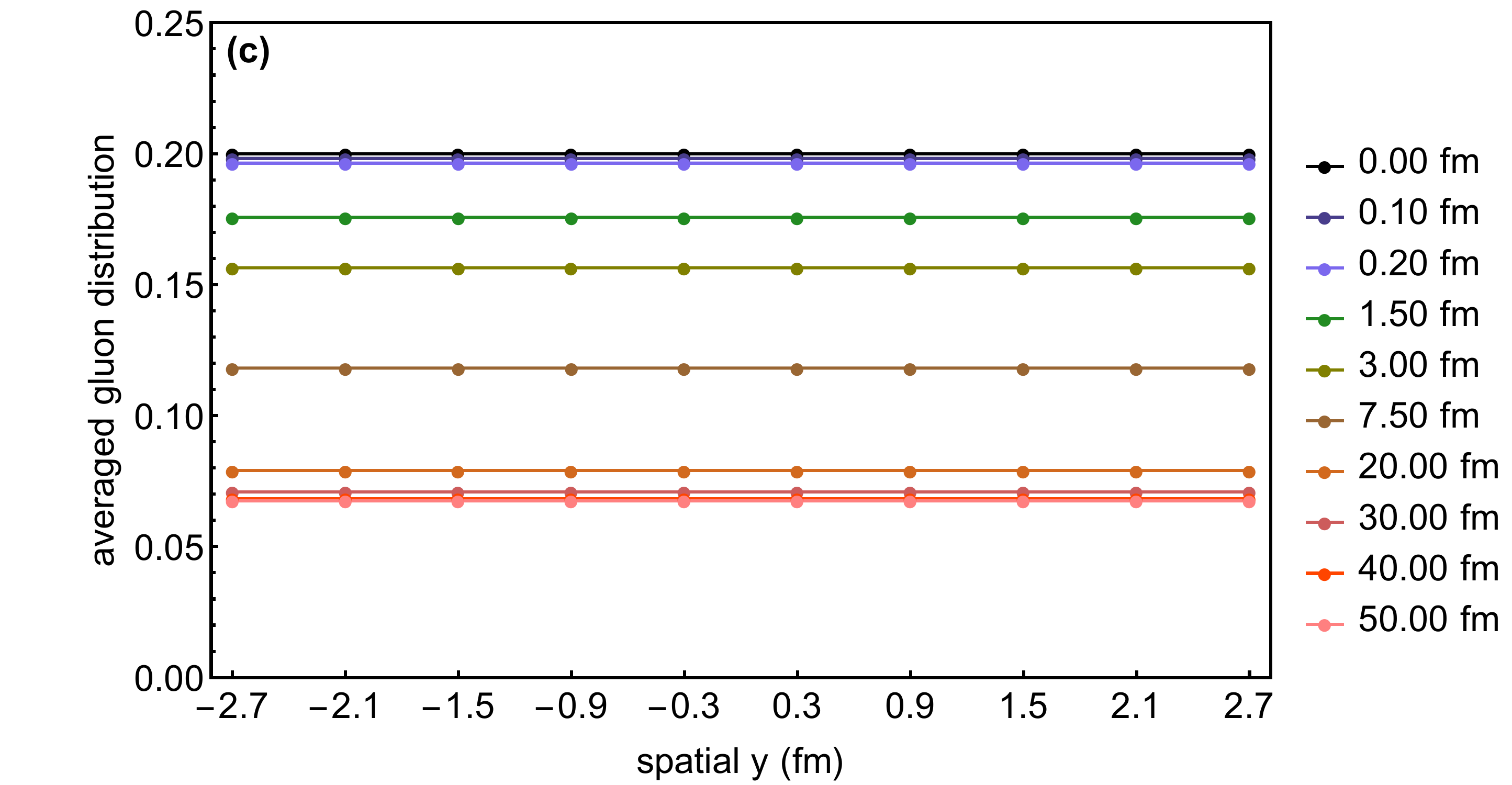}\includegraphics[scale=0.28]{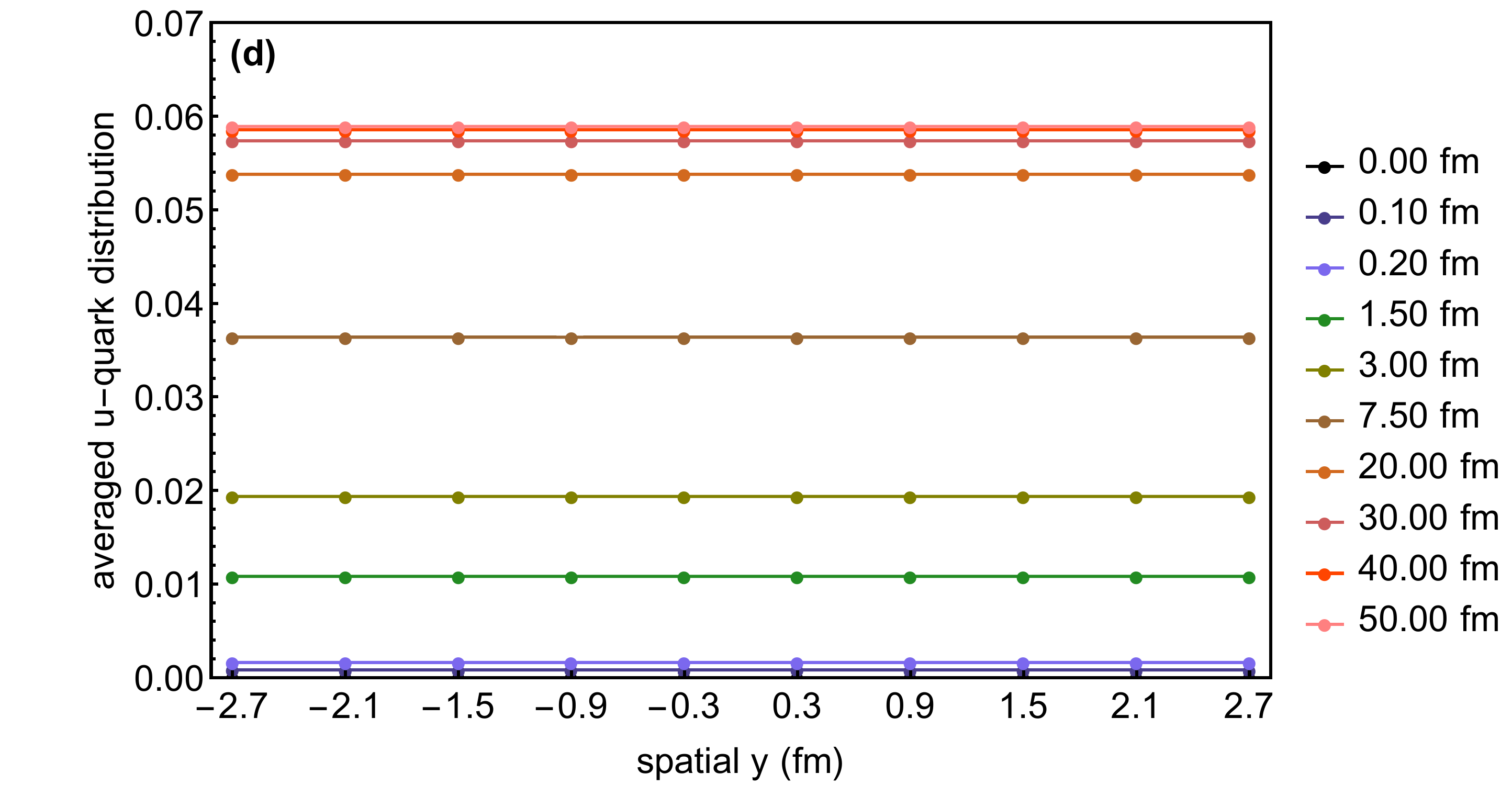}
\par\end{centering}
\centering{}\caption{\label{fig:Particle-distributions-in}
The gluon and u-quark distributions in spatial $x$ and $y$ directions at different times.
The phase space gird is
taken as $[n_{x},n_{y},n_{z},n_{px},n_{py},n_{pz}]=[10,10,1,10,10,1]$,
with $n$ being the number of grids. We have also chosen the coupling
constant $\alpha_{s}$ = 0.3, phase space box is of size $[-3\text{fm},3\text{fm}]^{3}\times[-2\text{GeV},2\text{GeV}]^{3}$and
$dt=0.0005\text{fm}$. The initial gluon distribution is
given by Eq. (\ref{eq:ini_f_test_spread}) with $|p_{x;max}|=2\text{GeV}$,
$|x_{x;max}|=3\text{fm}$, $f_{g;max}=0.2$ and $f_{g;min}=0.1$.
At each $x$ or $y$, we plot $f_{g}$ or $f_{u}$ which is averaged in $(y,z,\mathbf{p})$ or $(x,z,\mathbf{p})$, respectively.
The calculation takes 3312 seconds on one Nvidia Tesla V100  card.
}
\end{figure}

\begin{figure}
\begin{centering}
\includegraphics[scale=0.32]{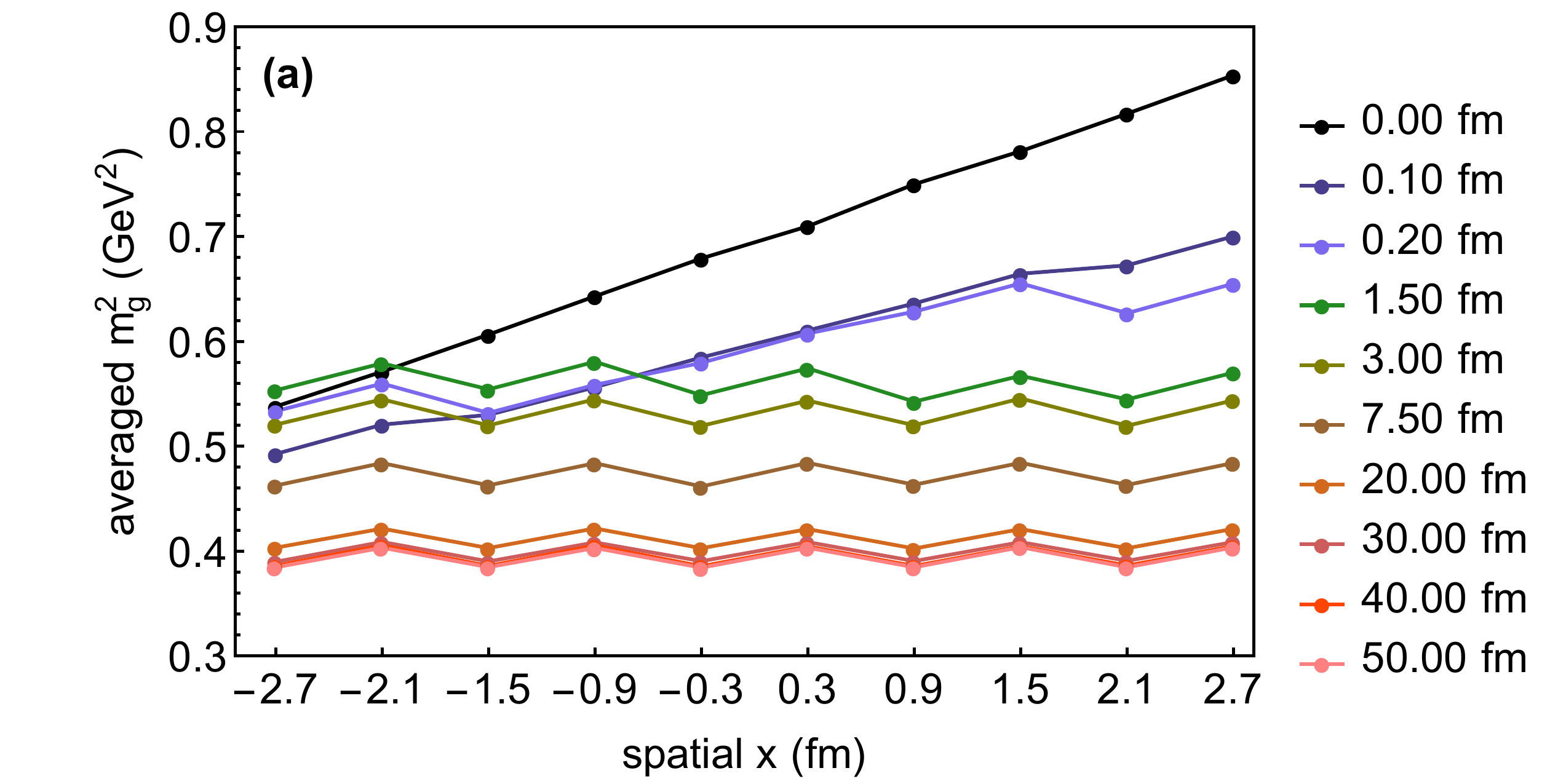}\includegraphics[scale=0.32]{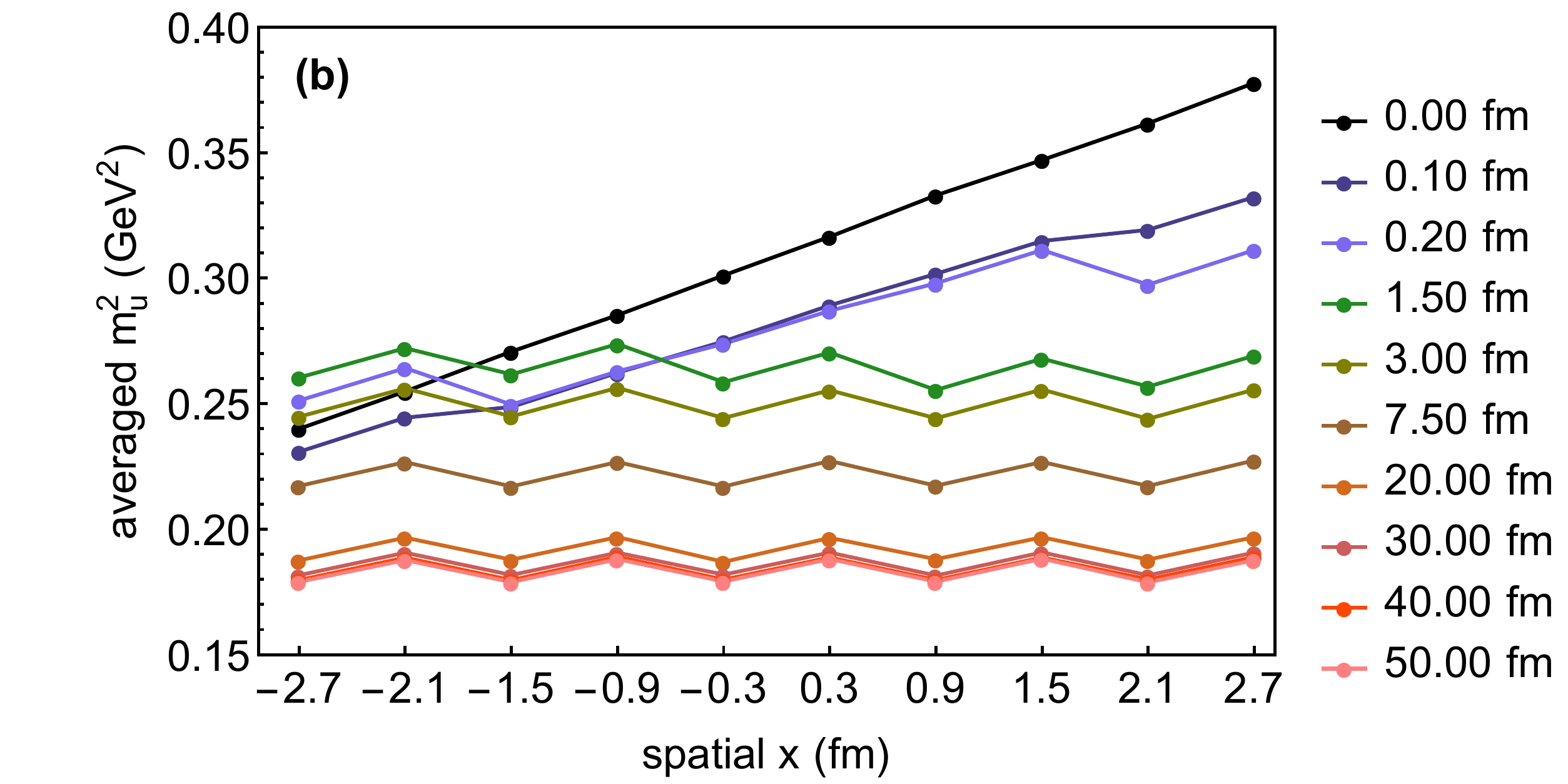}
\par\end{centering}
\begin{centering}
\includegraphics[scale=0.32]{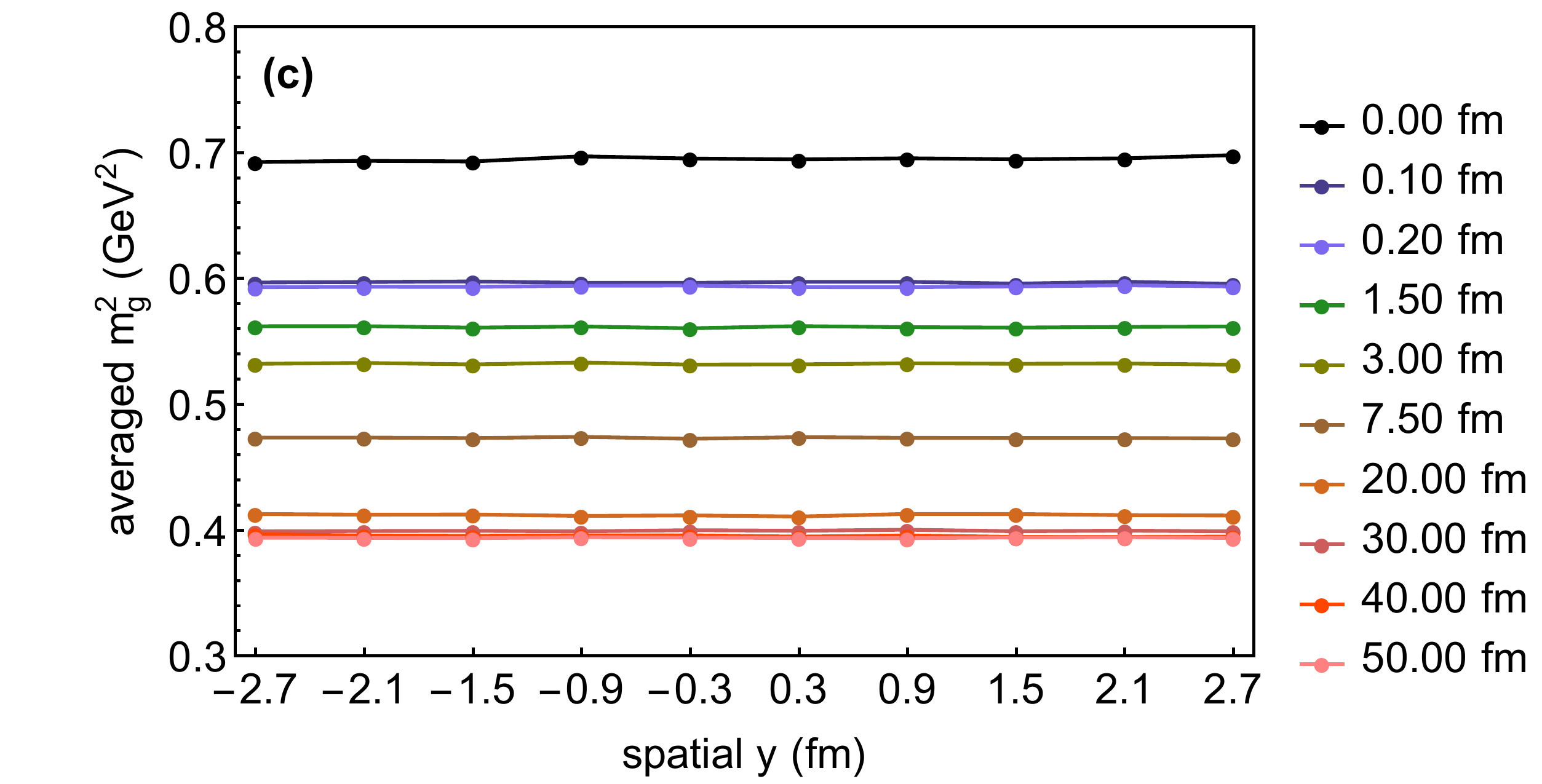}\includegraphics[scale=0.32]{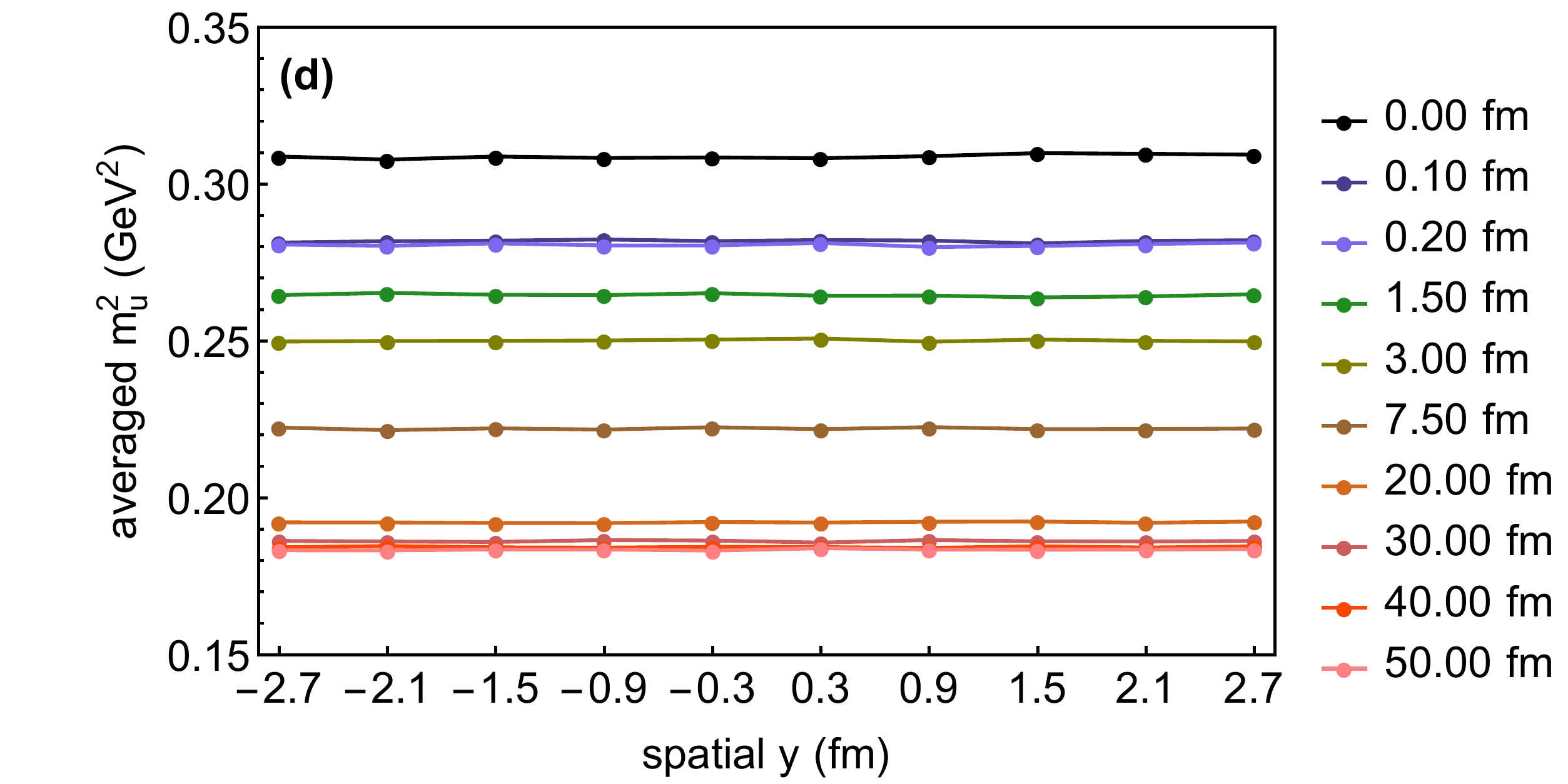}
\par\end{centering}
\centering{}\caption{\label{fig:Mass-distribution-in}
Distribution of mass squared $m_{g}^{2}$ and $m_{u}^{2}$ in spatial $x$ and $y$ directions at different times. The parameters
are chosen to be the same as in Fig \ref{fig:Particle-distributions-in}.}
\end{figure}

When the spatial part of the BE is taken into account, the evolution of the phase space distributions becomes more complicated.
Since the dynamical masses now also depend on the spatial grids as in Eq. (\ref{eq:mg}) and (\ref{eq:mq}), all differential terms in Eq. (\ref{eq:left handside}) contribute to the evolution.
The initial conditions for the distribution function must be physical, e.g. the fermion distribution functions should be smaller than unity, and all phase  distributions should be positive definite. Here for simplicity, we set the initial gluon distribution as follows:
\begin{eqnarray}
f_{g;\textrm{intial}}(\mathbf{x},\mathbf{p}) & = & \left[\frac{f_{g;max}-f_{g;min}}{2|x_{x;max}|}\right](x+|x_{x;max}|)+\left[-\frac{f_{g;max}-f_{g;min}}{2|p_{x;max}|}\right](p_{x}+|p_{x;max}|)\nonumber \\
 &  & +2f_{g;min},\label{eq:ini_f_test_spread}
\end{eqnarray}
where $f_{g;max}$ and $f_{g;min}$ are two parameters, which stand for the the maximum and minimum values of the distribution. $2|p_{x;max}|$ is the size of the momentum box in the $x$ direction, and $2|x_{x;max}|$ is the size of the spatial box in the $x$ direction.
One can also choose other types of initial conditions, and the main results will be similar.
The distribution function $f_{g}(\mathbf{x},\mathbf{p})$ in Eq. (\ref{eq:ini_f_test_spread}) linearly increases in the $x$ direction and linearly decreases in the $p_{x}$ direction.
For simplicity, we set the initial quark distribution to be zero,
\begin{equation}
f_{q(\bar{q});\textrm{initial}}=0.\label{eq:ini_f_test_spread_quark}
\end{equation}

In Fig. \ref{fig:Particle-distributions-in}, we show the gluon and $u$-quark distributions as a function of the spatial directions
$x$ and $y$ at different evolution times.
Initially, the gluon distribution is linear in spatial $x$ and $u$ quark distribution is vanishing.
As the time evolves, the gluons tend to convert to quarks, similar to Fig. \ref{fig:Comparison-between-with-2}.
In the end, the distributions of both $u$ quarks and gluons become approximately uniform in all spatial directions (Here we show the distribution functions in the $x$ and $y$-directions). Other fermions have a similar pattern as well.

In Fig. \ref{fig:Mass-distribution-in}, we show the dynamical masses for gluons and $u$ quarks in the spatial $x$ and $y$-directions
at different evolution times. Initially, both masses squared $m_{g}^{2}$ and $m_{u}^{2}$ increase linearly with $x$. Then  as time evolves, they tend to become homogenous in spatial $x$ direction. The distributions for other flavors of quarks and anti-quarks are similar.

\subsection{Evolution of distribution functions in momentum space and gluon condensation \label{subsec:Evolution-in-momentum}}

\begin{figure}
\begin{centering}
$\ $\includegraphics[scale=0.5]{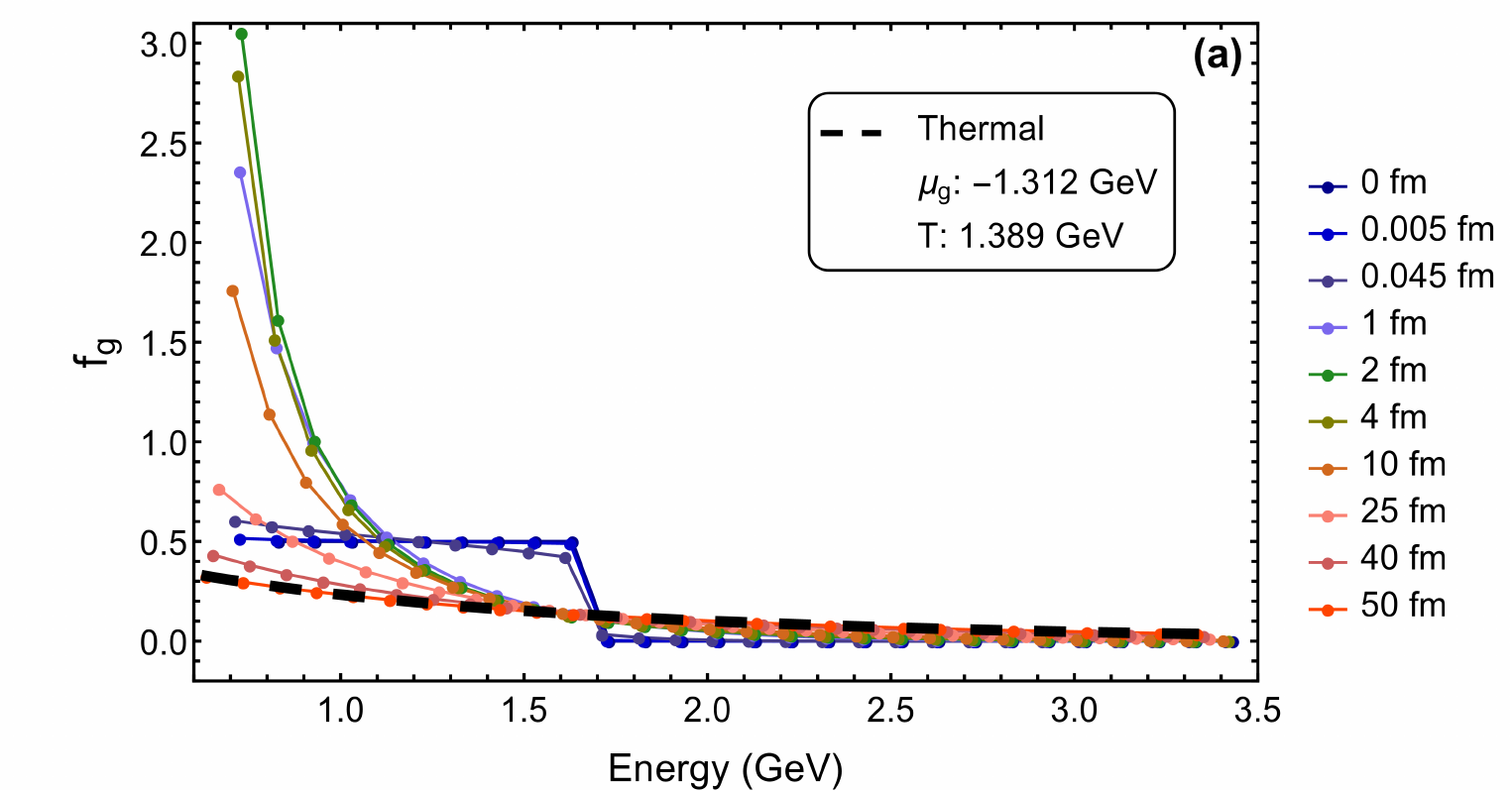}\includegraphics[scale=0.5]{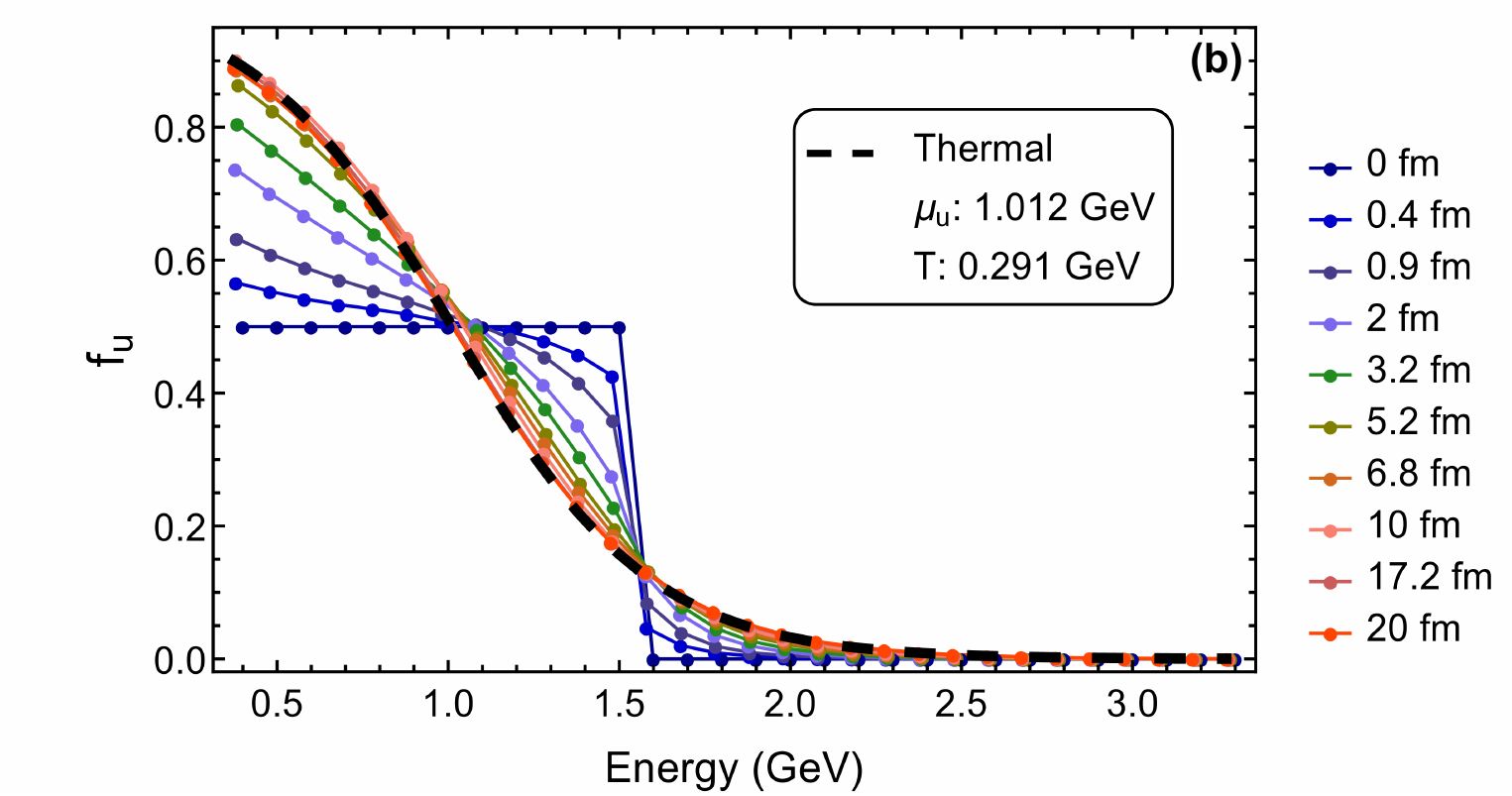}
\par\end{centering}
\begin{centering}
\includegraphics[scale=0.5]{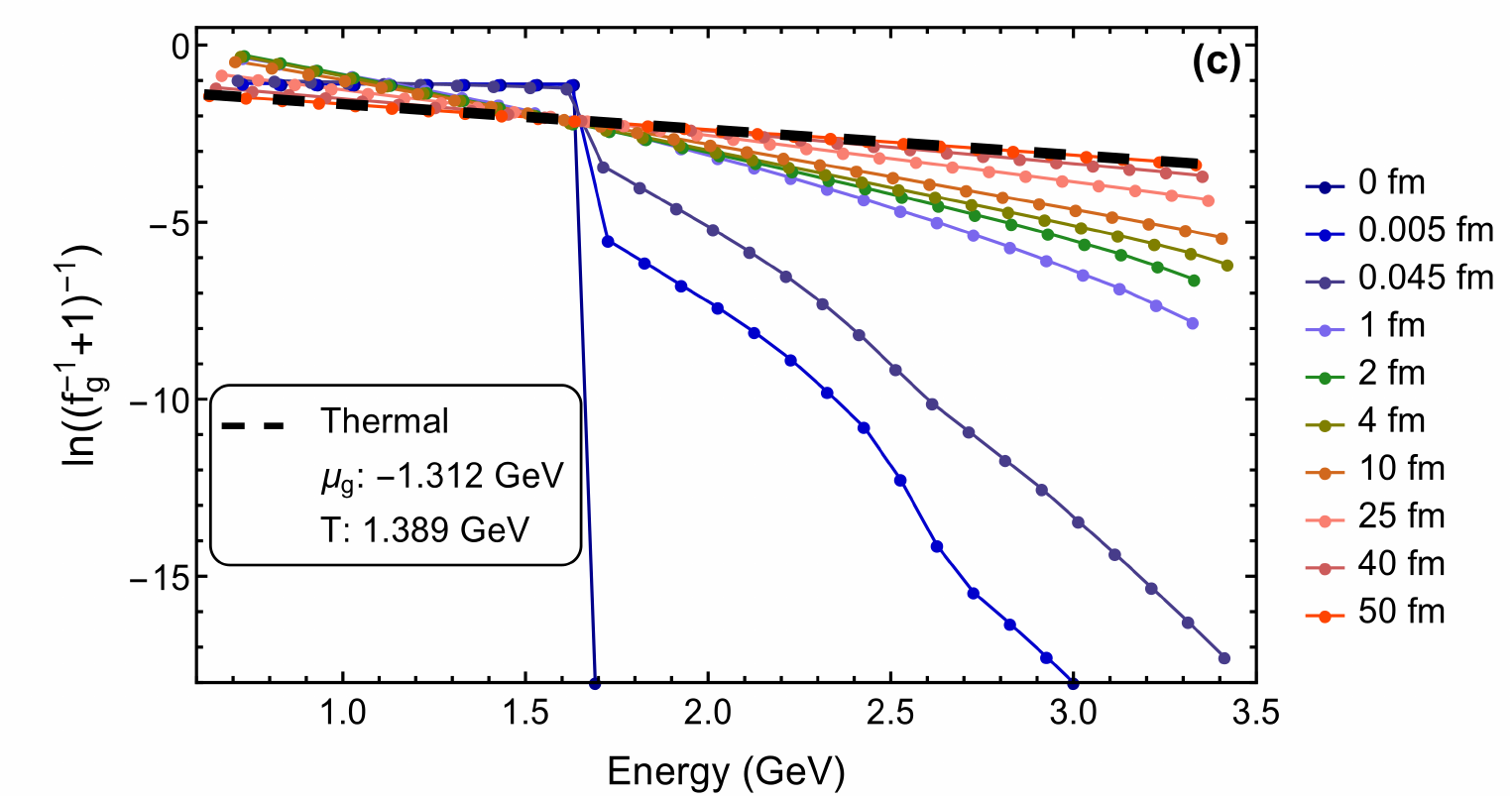}\includegraphics[scale=0.5]{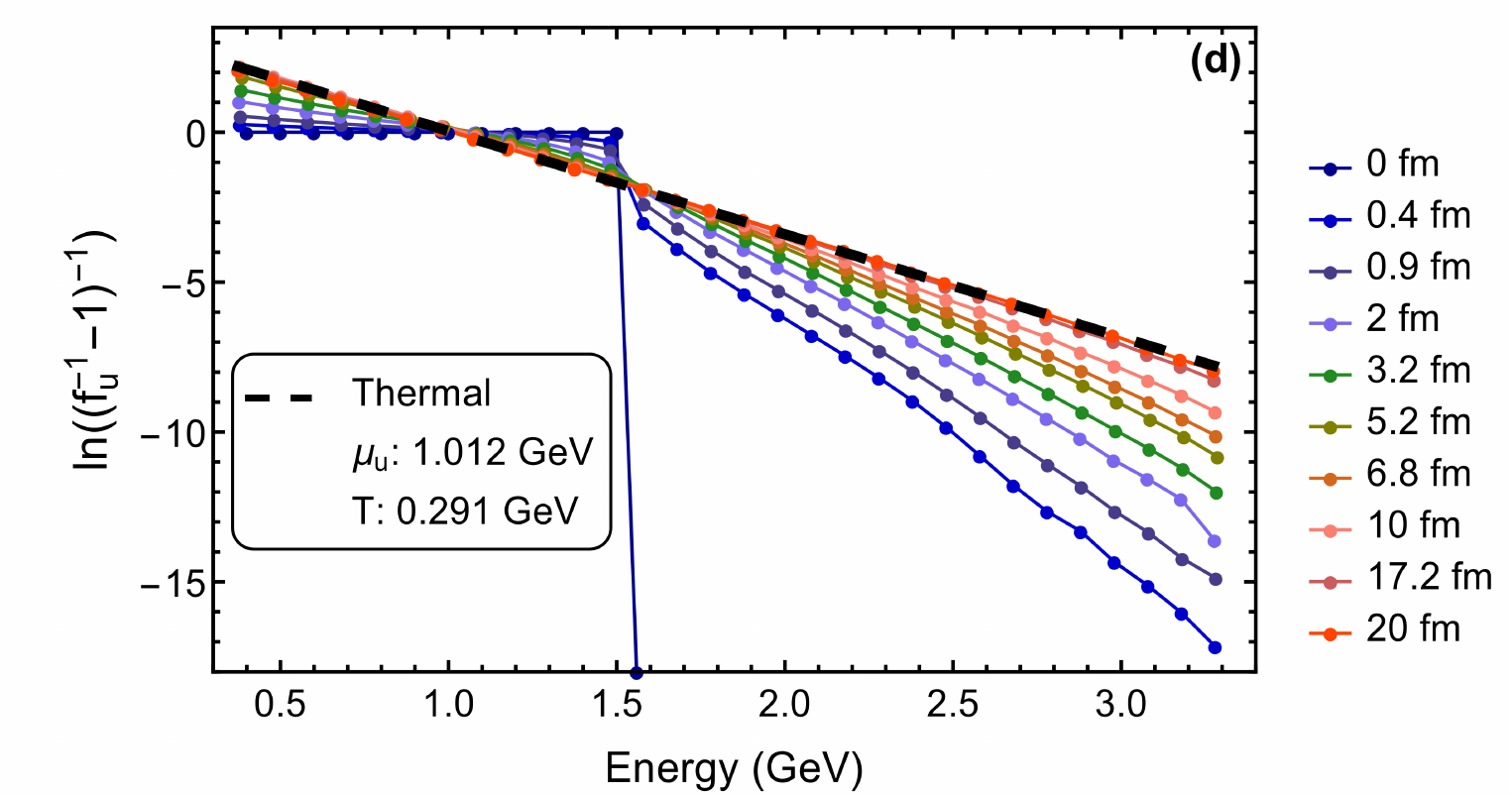}
\par\end{centering}
\centering{}\caption{\label{fig:Evolution-from-CGC}
Time evolution of the parton distribution functions from initial step functions to thermal distributions.
Panels (a) and (c) show the results for pure gluons while panels (b) and (d) for pure quarks.
The phase space gird is taken as $[n_{x},n_{y},n_{z},n_{px},n_{py},n_{pz}]=[1,1,1,30,30,30]$.
The coupling constant $\alpha_{s}$ = 0.3, the phase space box is of size $[-3\text{fm},3\text{fm}]^{3}\times[-2\text{GeV},2\text{GeV}]^{3}$, with $f_{0}=0.5$ and $Q_{s}=1.5\text{GeV}$.
For pure gluons, the time step is taken as $dt=0.00005\text{fm}$ and for pure quarks $dt=0.001\text{fm}$.
On one Nvidia Tesla V100 card, the evaluation from 0 fm to 50 fm takes 60 hours for pure gluon case and 2 hours for pure fermions case from 0 fm to 20 fm.}
\end{figure}

\begin{figure}
\begin{centering}
\includegraphics[scale=0.38]{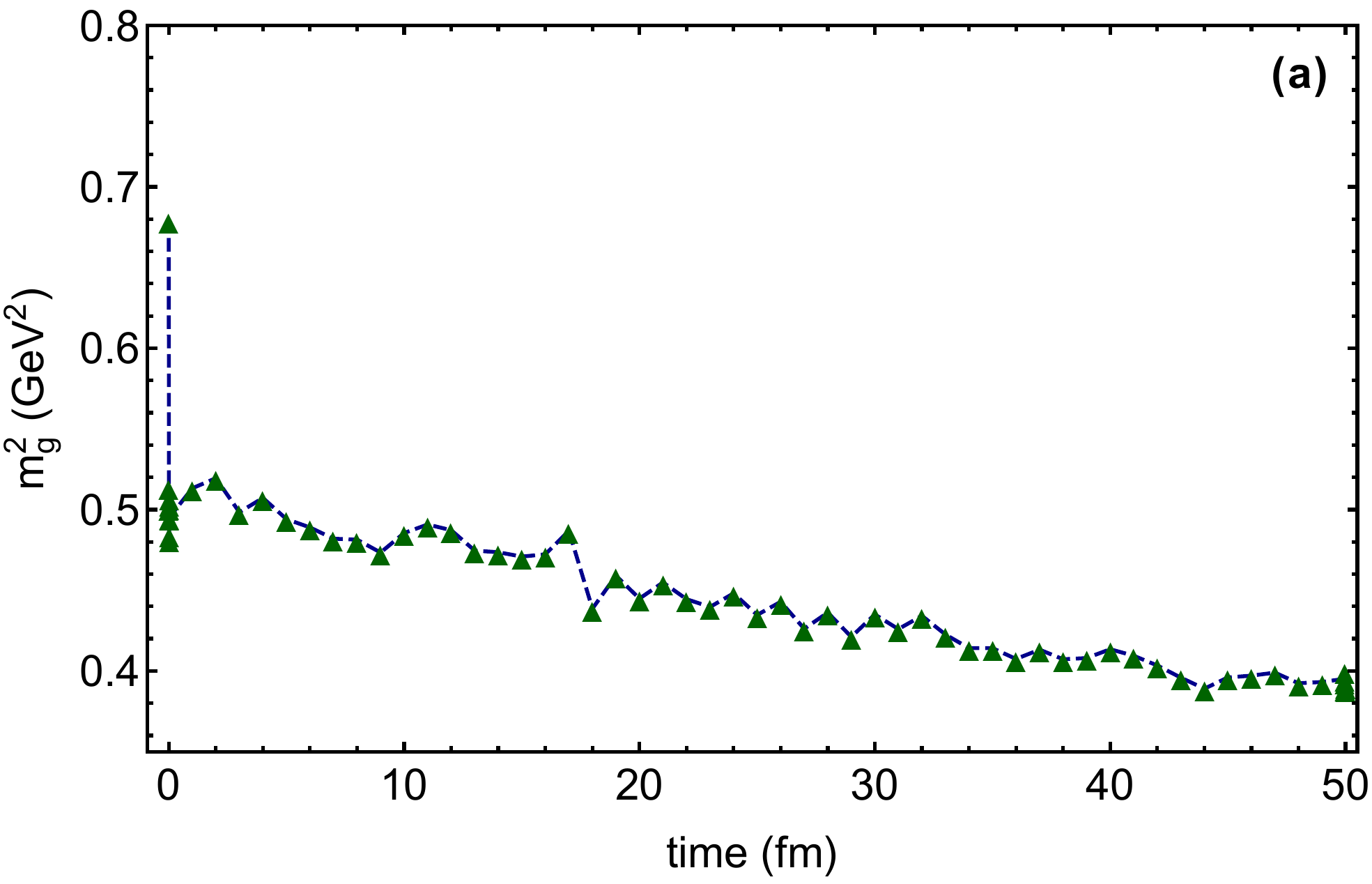}\includegraphics[scale=0.38]{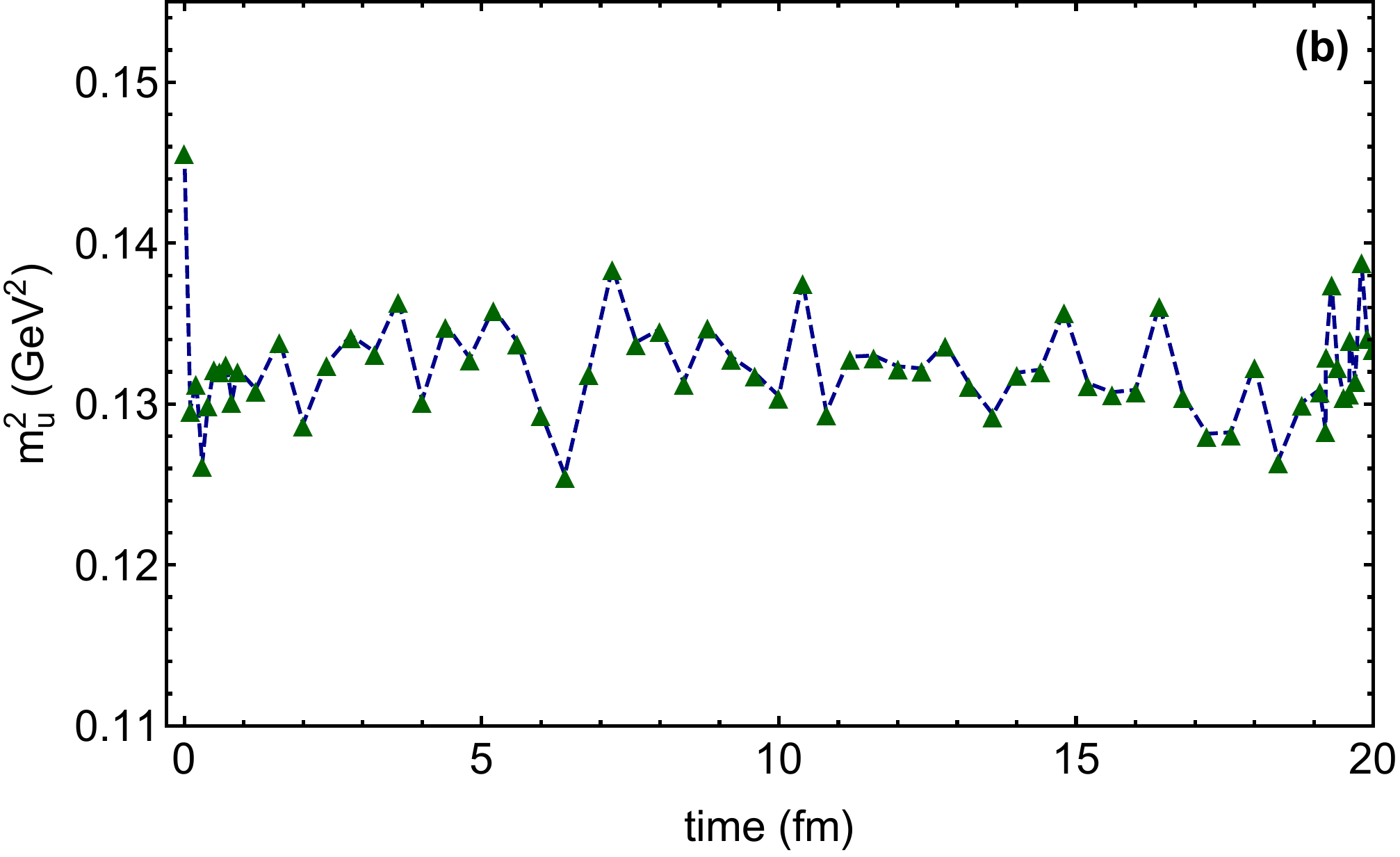}
\par\end{centering}
\caption{\label{fig:Evolution-of-fermions}
Time evolution of masses squared for pure gluons and pure quarks from initial step functions to thermal distributions.
Panels (a) and (b) show the results for gluons and quarks, respectively.
The parameters are chosen to be the same as in Fig. \ref{fig:Evolution-from-CGC}.}
\end{figure}

Now we show the evolution of systems in momentum space. For simplicity, we neglect the spatial dependence and set the systems to be homogenous in coordinate space. As a first attempt, we investigate the time evolution of pure gluon or pure quark systems in momentum space. The initial gluon distribution is chosen to be a step function,
\begin{eqnarray}
f_{g}(\mathbf{p}) & = & f_{g,0}\theta\left(1-\frac{|\mathbf{p}|}{Q_{s}}\right),\label{eq:CGC}
\end{eqnarray}
where $f_{0}$ and $Q_{s}$ are parameters.
This initial condition is to mimic the distribution from Color Glass Condensation (CGC) \citep{McLerran2008}, with $Q_{s}$ being the saturation scale.
The evolution of the distribution function for a pure gluon system is shown in Fig. \ref{fig:Evolution-from-CGC} (a) and (c). Note that for gluon evolution, the time $dt$ is set very small (in our case $dt=0.00005\text{fm}$) to ensure that the distribution functions are positive.
Similarly for a pure quark system, we choose the initial quark distribution function as,
\begin{equation}
f_{q}(\mathbf{p})=f_{q,0}\theta\left(1-\frac{|\mathbf{p}|}{Q_{s}}\right).\label{eq:CGC_quark}
\end{equation}
The time evolution of the pure $u$ quark distribution function is shown in Fig. \ref{fig:Evolution-from-CGC} (b) and (d). For pure quark
system, the time step $dt$ can be relative larger (in our case $dt=0.001\text{fm}$).

From Fig. \ref{fig:Evolution-from-CGC}, we find that the thermalization of gluons is quite different from that of quarks.
The gluons will first accumulate in the soft region, where the energy is smaller than $1.0\textrm{GeV}$.
This phenomenon may indicate the gluon condensation.
While for quarks, we have not observed such phenomenon.
The gluon condensation may be of importance to the pre-thermalization of quark-gluon plasma created in the heavy-ion collisions.
We will investigate such issue in our future studies based on our program.
It is noted that at the thermal equilibrium, the gluon chemical potential is negative for pure gluon system while the quark chemical potential is positive for pure quark system in Fig. \ref{fig:Evolution-from-CGC}.

In Fig. \ref{fig:Evolution-of-fermions}, we show the evolution of the dynamical masses squared for gluons and quarks.
For a pure gluon system, the gluon's thermal mass decreases with time and eventually becomes stable as the system reaches thermal equilibrium.
For a pure quark system, the quark's thermal mass oscillates with time, but the averaged value tends to be constant.

\begin{figure}
\begin{centering}
\includegraphics[scale=0.52]{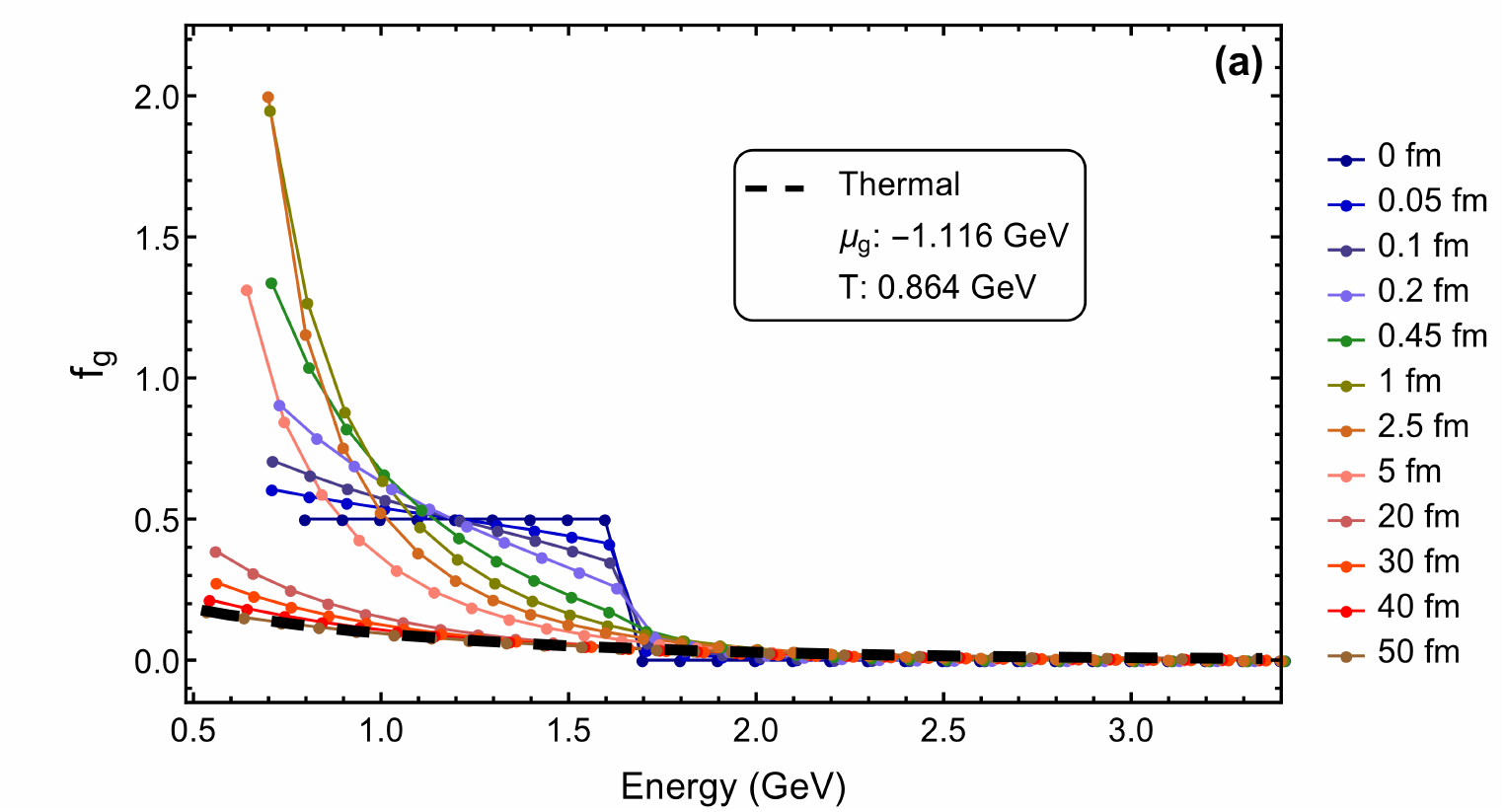}\includegraphics[scale=0.52]{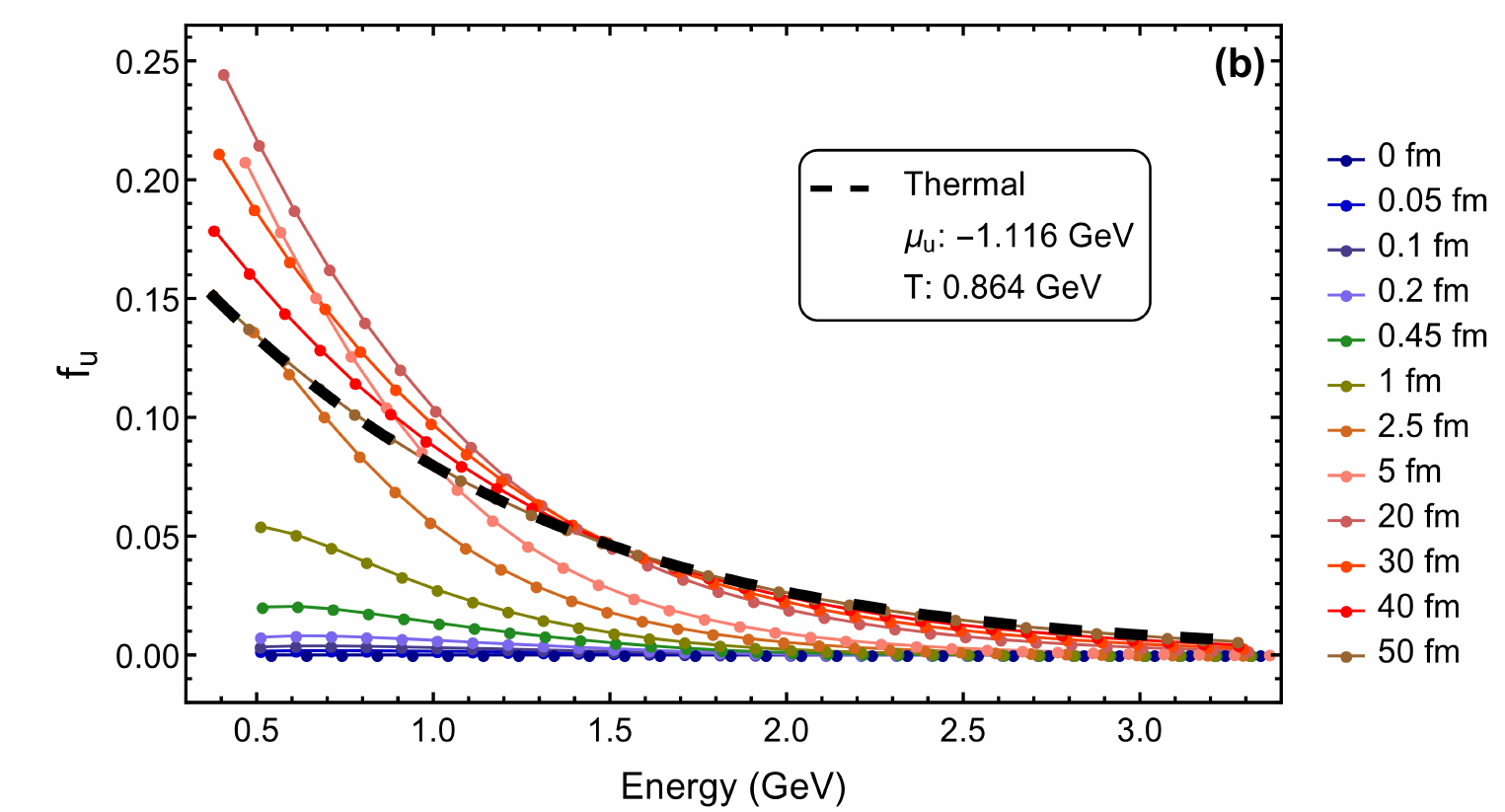}
\par\end{centering}
\begin{centering}
\includegraphics[scale=0.52]{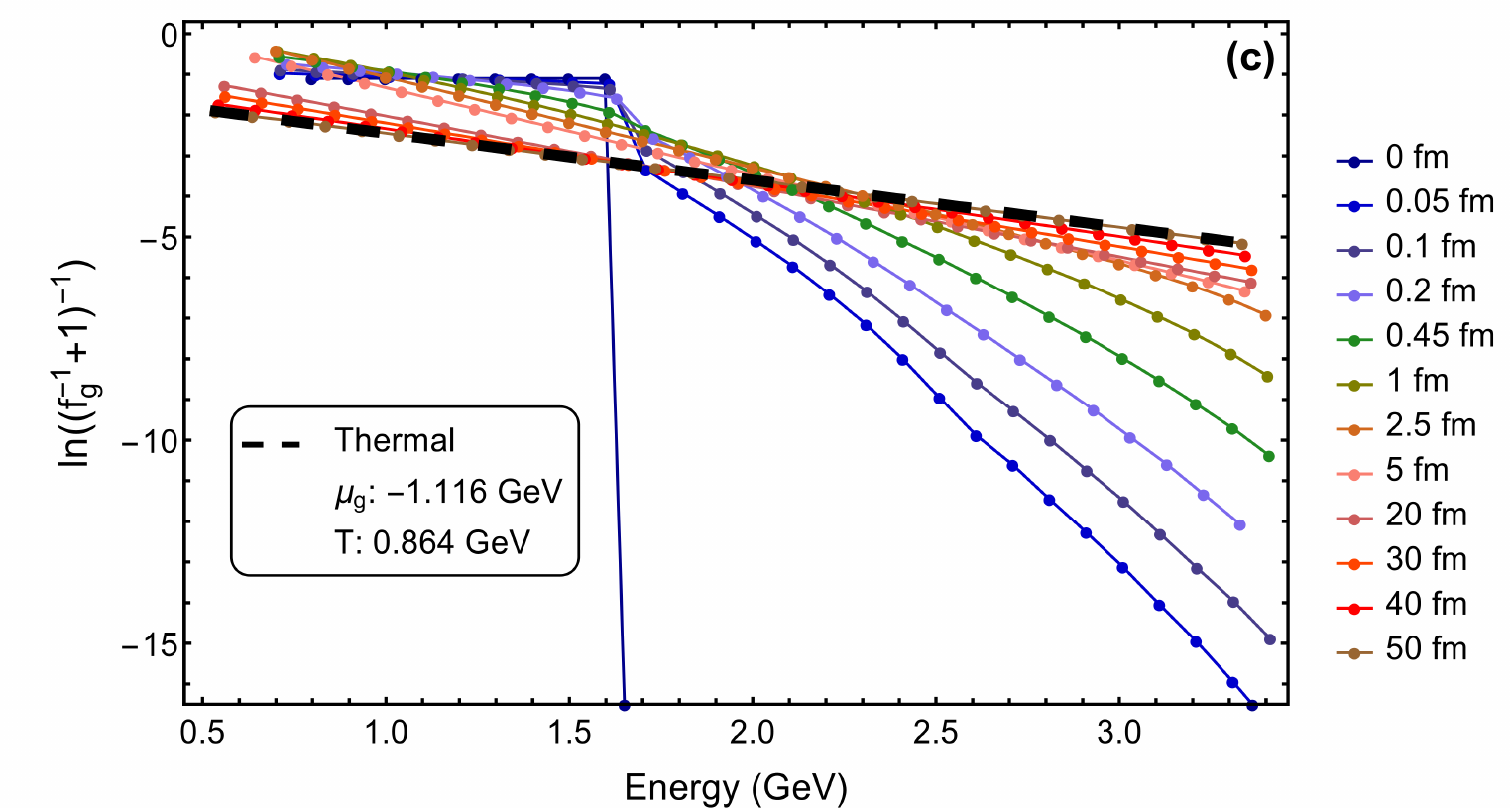}\includegraphics[scale=0.52]{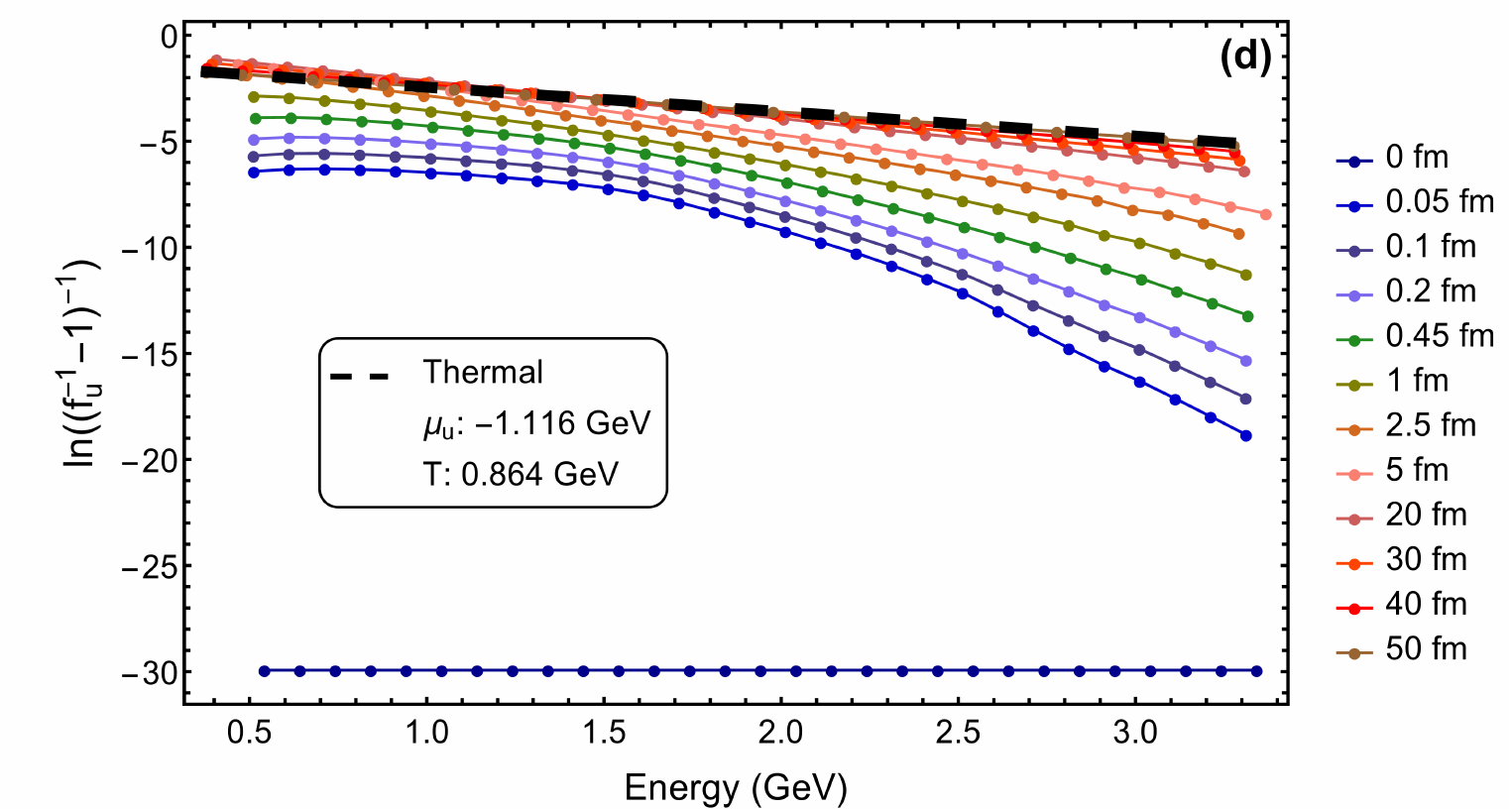}
\par\end{centering}
\caption{\label{fig:Evolution-of-systems}
Time evolution of the parton distribution functions for quark-gluon systems.
The initial condition is $f_{q}(\mathbf{p})=0$ for quarks and $f_{g}(\mathbf{p})=f_{g,0}\theta(1-|\mathbf{p}|/Q_{s})$ for gluons, with $f_{0}=0.5$ and $Q_{s}=1.5\text{GeV}$.
The phase space gird is taken as $[n_{x},n_{y},n_{z},n_{px},n_{py},n_{pz}]=[1,1,1,30,30,30]$.
The coupling $\alpha_{s}$ = 0.3, the phase space box is of size $[-3\text{fm},3\text{fm}]^{3}\times[-2\text{GeV},2\text{GeV}]^{3}$, and the time step is taken as $dt=0.00005\textrm{fm}$.
On one Nvidia Tesla V100 card, the evaluation from 0 fm to 50 fm takes around 2 days.}
\end{figure}

\begin{figure}
\begin{centering}
\includegraphics[scale=0.38]{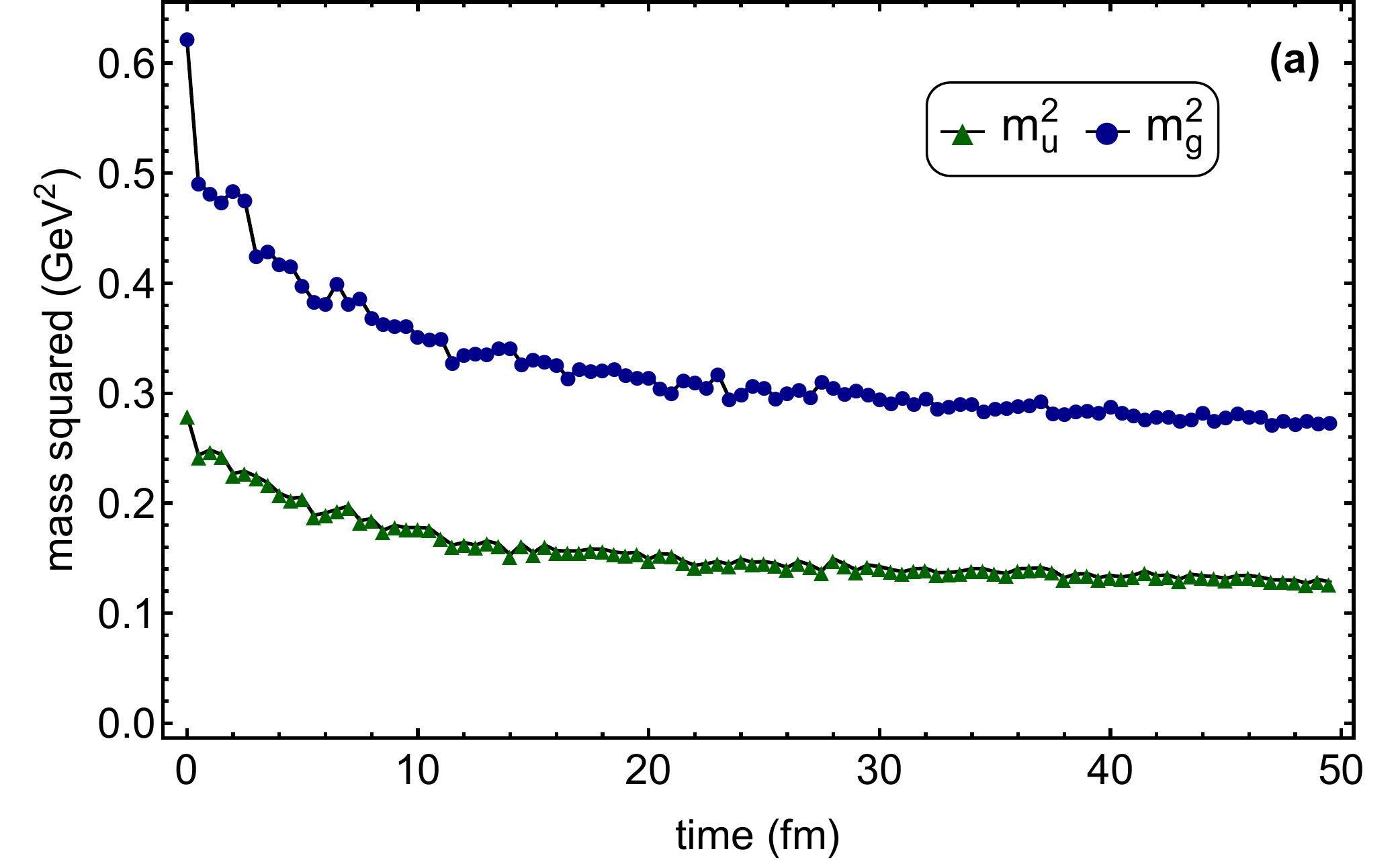}\includegraphics[scale=0.38]{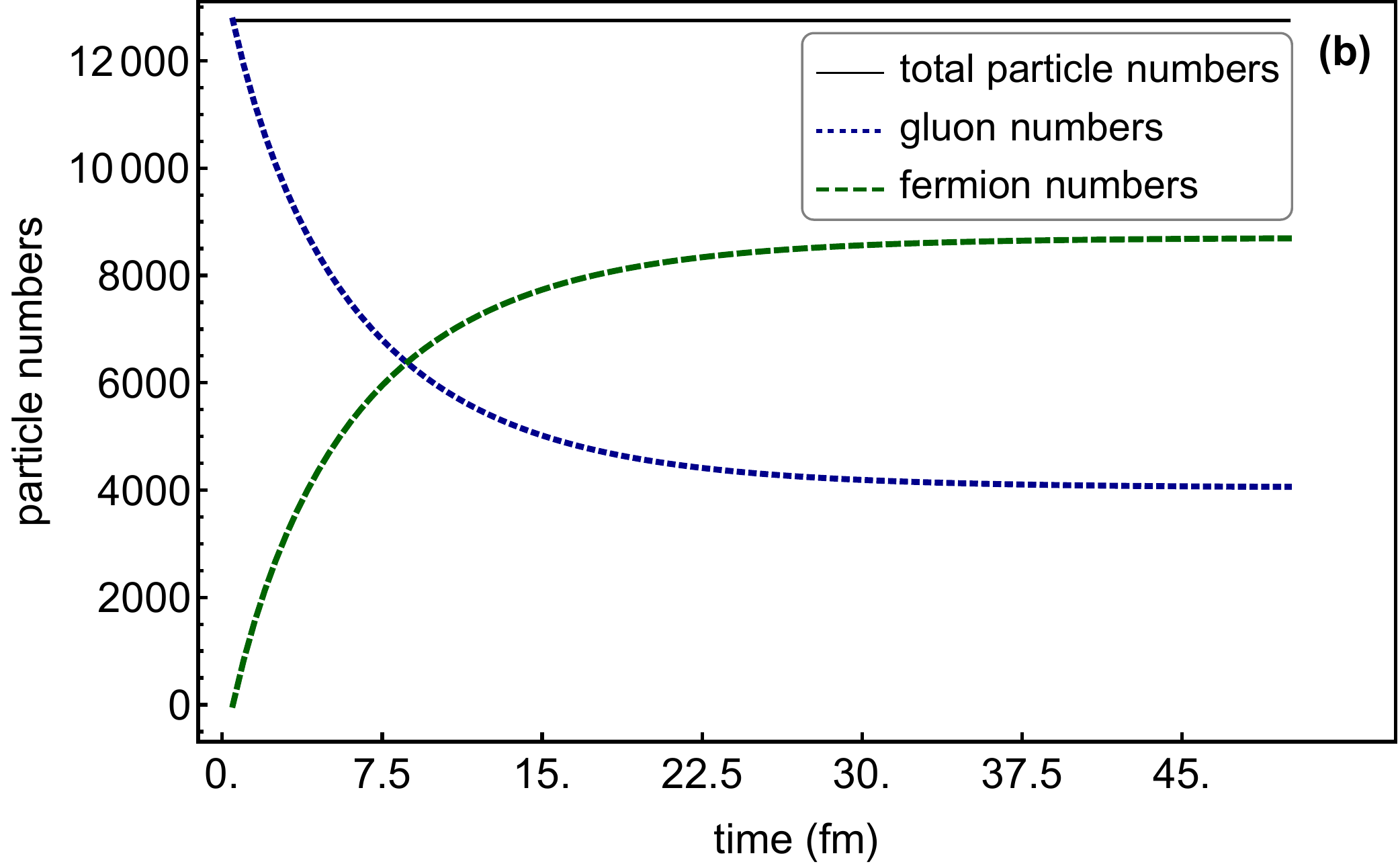}
\par\end{centering}
\caption{\label{with_quark_gluon_mass}
Time evolution of masses squared $m_{g,u}^{2}$ and the particle numbers for quark-gluon systems.
The initial condition and parameters are take as the same in Fig. \ref{fig:Evolution-of-systems}.}
\end{figure}

We now extend our simulation to a system composed of both quarks and gluons.
The initial condition for gluons is set the same in Eq. (\ref{eq:CGC}), and for quarks we choose $f_{q}(\mathbf{p})=0$.
In Fig. \ref{fig:Evolution-of-systems}, we show the distribution functions of gluons and $u$ quarks as a function of parton energy at different evolution times.
For gluons, the distribution function in the soft region ($\lesssim1.0\textrm{GeV}$) increases very fast at the beginning and then decreases to the thermal equilibration.
The accumulation of gluons in the soft region implies possible gluon condensation.
For quarks, the distribution function reaches the Fermi-Dirac distribution gradually.
We also notice that the chemical potentials for gluons and quarks are both negative in the thermal equilibrium.
Compared with the result in Fig. \ref{fig:Evolution-from-CGC} (b) and (d), the negative chemical potential for quarks might come from different initial condition and their interaction with gluons.

In Fig. \ref{with_quark_gluon_mass}, we show the dynamical masses squared $m_{g,u}^{2}$ and the particle numbers as a function of the evolution time.
It is interesting that there is no oscillation behavior for $m_{u}^{2}$ here, as opposed to the results in Fig. \ref{fig:Evolution-of-fermions}.
Instead, $m_{u}^{2}$ decreases with time and then reaches to a constant, similarly to $m_{g}^{2}$.
In Fig. \ref{with_quark_gluon_mass} (b), we also confirm that the total particle number is strictly conserved.
The gluons convert to quarks in a very short time ($\lesssim10\textrm{fm}$), and then both gluon and quark numbers tend to be constant.

\section{Conclusion and discussion \label{sec:Conclusion}}

In this work, we have developed a new numerical framework for obtaining the full solutions of relativistic BE on GPUs.
Our main equation, i.e., the complete relativistic BE, is of form Eq. (\ref{eq:boltzmann equation general form}).
We have considered the thermal systems of $3$ flavor quarks, anti-quarks and gluons.
For simplicity, we only consider $2\rightarrow2$ scattering processes, in which the total particle numbers are conserved.
Since the quarks and gluons have dynamical masses in Eqs. (\ref{eq:mg}, \ref{eq:mq}), there is an external force in Eq. (\ref{eq:boltzmann equation general form}).
Also the kinetic energy-momentum tensor in Eq. (\ref{eq:current conservation}) is not conserved due to the external force.

To solve BE numerically, we first rewrite the main equation as Eq. (\ref{eq:left handside}) with its discrete form in Eq. (\ref{eq:descrete left hand}).
There are two ways to handle the $\delta$-function in the collisional integral, either integrating out $\mathbf{k}_{2}$ or integrating out $\mathbf{k}_{3}$.
In this work, we have chosen the first choice in Eq. (\ref{eq:work out delta E}), which is more stable than the second one as shown in Fig. \ref{fig:Comparison-of-the}.
Next we introduce the ``symmetrical sampling'' method to ensure the conservation of the total particle number.
We have also investigated the energy conservation which is not strictly conserved numerically due to the discrete grids.
However, the numerical errors will decrease very fast if we increase the numbers of the grids, and the conservation of the total energy can be achieved up to 99.8\% in our calculation.

We have studied the time evolution of the distribution functions in both coordinate and momentum spaces.
Fig. \ref{fig:Particle-distributions-in} has shown the gluon and $u$-quark distributions as a function of spatial direction $x$ at different evolution time, given the initial condition in Eqs. (\ref{eq:ini_f_test_spread}, \ref{eq:ini_f_test_spread_quark}).
It is found that the distributions of both $u$ quarks and gluons become homogeneous in the coordinate space at a later time, implying the reach of thermal equilibrium.
Fig. \ref{fig:Evolution-from-CGC} and Fig. \ref{fig:Evolution-of-systems} show the evolution of the distribution functions in the momentum space for pure gluons, pure quarks and quark-gluon mixtures. It is interesting that the thermalization process of gluons is different from that of quarks.
The fermions reach the thermal distribution smoothly, while the distribution functions of gluons at an early time increase very fast in the soft region and then decrease to thermal distributions.

In summary, we have provided a full numerical solution to the BE with complete $2\rightarrow2$ scattering processes with high computing performance.
Our framework may serve as a basis to study the pre-thermalization stage in heavy-ion collisions in the future.
Currently, our program can use the grid sizes up to $n_{x},n_{y},n_{z}=20$ and $n_{px},n_{py},n_{pz}=40$.
Very large grid sizes such as $n_{x},n_{y},n_{z},n_{px},n_{py},n_{pz}\geq50$ are still challenging, even with the help of GPU clusters.
We will continue to improve our framework and algorithms along this direction.
In the future, we will include $2\rightarrow3$ processes and study the interplay between elastic and inelastic scatterings.
We may also include the external electromagnetic fields to study the quantum transport phenomena under the strong electromagnetic fields.

\section*{acknowledgments}

The authors thank Jean Paul Blaizot, Aleksas Mazeliauskas, Xin-nian Wang, Shuai Liu and Ren-hong Fang for useful comments and discussions. S.~P. and Q.~W. are grateful to the Yukawa Institute for Theoretical Physics (YITP) for its hospitality.
This work is supported in part by the Natural Science Foundation of China (NSFC) under Grants No. 11535012 and 11890713.
S.P. is supported by One Thousand Talent Program for Young Scholars.
G.-Y. Q. is supported by NSFC under Grants No. 11775095, 11890711, 11375072 and by the China Scholarship Council (CSC) under Grant No. 201906775042.

\section*{Appendix}

\subsection{Matrix elements squared for $2\rightarrow2$ scattering \label{matrix element}}

\begin{table}[H]
\centering{}
\begin{tabular}{c|c}
\hline
$ab\rightarrow cd$ & $\left\vert M_{a\left(k_{1}\right)b\left(k_{2}\right)\rightarrow c\left(k_{3}\right)d\left(p\right)}\right\vert ^{2}$\tabularnewline
\hline
$\begin{array}{c}
q_{1}q_{2}\rightarrow q_{1}q_{2}\\
\bar{q}_{1}q_{2}\rightarrow\bar{q}_{1}q_{2}\\
q_{1}\bar{q}_{2}\rightarrow q_{1}\bar{q}_{2}\\
\bar{q}_{1}\bar{q}_{2}\rightarrow\bar{q}_{1}\bar{q}_{2}
\end{array}$ & $8g^{4}\dfrac{d_{F}^{2}C_{F}^{2}}{d_{A}}\left(\dfrac{s^{2}+u^{2}}{(t-m_{g}^{2})^{2}}\right)$\tabularnewline
\hline
$\begin{array}{c}
q_{1}q_{1}\rightarrow q_{1}q_{1}\\
\bar{q}_{1}\bar{q}_{1}\rightarrow\bar{q}_{1}\bar{q}_{1}\\
\\
\end{array}$ & $8g^{4}\dfrac{d_{F}^{2}C_{F}^{2}}{d_{A}}\left(\dfrac{s^{2}+u^{2}}{(t-m_{g}^{2})^{2}}+\dfrac{s^{2}+t^{2}}{(u-m_{g}^{2})^{2}}\right)$$+16g^{4}d_{F}C_{F}\left(C_{F}-\frac{C_{A}}{2}\right)\dfrac{s^{2}}{(t-m_{g}^{2})(u-m_{g}^{2})}$\tabularnewline
\hline
$\begin{array}{c}
q_{1}\bar{q}_{1}\rightarrow q_{1}\bar{q}_{1}\\
\\
\end{array}$ & $8g^{4}\dfrac{d_{F}^{2}C_{F}^{2}}{d_{A}}\left(\dfrac{s^{2}+u^{2}}{(t-m_{g}^{2})^{2}}+\dfrac{u^{2}+t^{2}}{s^{2}}\right)$$+16g^{4}d_{F}C_{F}\left(C_{F}-\frac{C_{A}}{2}\right)\dfrac{u^{2}}{(t-m_{g}^{2})s}$\tabularnewline
\hline
$\begin{array}{c}
q_{1}\bar{q}_{1}\rightarrow q_{2}\bar{q}_{2}\\
\\
\end{array}$ & $8g^{4}\dfrac{d_{F}^{2}C_{F}^{2}}{d_{A}}\dfrac{t^{2}+u^{2}}{s^{2}}$\tabularnewline
\hline
$\begin{array}{c}
q_{1}\bar{q}_{1}\rightarrow gg\\
\\
\end{array}$ & $8g^{4}d_{F}C_{F}^{2}\left(\dfrac{u}{(t-m_{g}^{2})}+\dfrac{t}{(u-m_{g}^{2})}\right)-8g^{4}d_{F}C_{F}C_{A}\left(\dfrac{t^{2}+u^{2}}{s^{2}}\right)$\tabularnewline
\hline
$\begin{array}{c}
q_{1}g\rightarrow q_{1}g\\
\bar{q}_{1}g\rightarrow\bar{q}_{1}g
\end{array}$ & $-8g^{4}d_{F}C_{F}^{2}\left(\dfrac{u}{s}+\dfrac{s}{(u-m_{g}^{2})}\right)+8g^{4}d_{F}C_{F}C_{A}\left(\dfrac{s^{2}+u^{2}}{(t-m_{g}^{2})^{2}}\right)$\tabularnewline
\hline
$\begin{array}{c}
gg\rightarrow gg\\
\\
\end{array}$ & $16g^{4}d_{A}C_{A}^{2}\left(3-\dfrac{su}{(t-m_{g}^{2})^{2}}-\dfrac{st}{(u-m_{g}^{2})^{2}}-\dfrac{tu}{s^{2}}\right)$\tabularnewline
\hline
\end{tabular}
\caption{\label{tab:Matrix-elements-squared} Matrix elements squared for all $2\rightarrow2$ parton scattering processes in QCD. The helicities and colors of all initial and final state particles are summed over.
$q_{1}$ ($\bar{q}_{1}$) and $q_{2}$ ($\bar{q}_{2}$) represent quarks (antiquarks) of different flavors, and $g$ represents the gluon. $d_{F}$ and $d_{A}$ denote the dimensions of the fundamental and adjoint representations of $SU_{c}(N)$ gauge group while $C_{F}$ and $C_{A}$ are the corresponding quadratic Casimirs.
In a $SU_{c}(3)$ theory with fundamental representation fermions, $d_{F}=C_{A}=3$, $C_{F}=4/3$, and $d_{A}=8$.
The infrared divergence is suppressed by introducing a regulator in the denominator \citep{Zhang1998a,Arnold2003a,Chen2013}.}
\end{table}

\bibliographystyle{h-physrev}
\bibliography{BEGPU}

\end{document}